\documentclass[12pt]{article}
\usepackage{amssymb}
\usepackage{color}
\usepackage{graphicx}

\newcommand{\mb}{\mathbb}
\newcommand{\A}{{\mathfrak A}}

\newcommand{\R}{{\cal R}}
\newcommand{\Sc}{{\cal S}}

\newcommand{\mc}{\mathcal}

\newcommand{\be}{\begin{equation}}
\newcommand{\en}{\end{equation}}
\newcommand{\bea}{\begin{eqnarray}}
\newcommand{\ena}{\end{eqnarray}}

\newcommand{\id}{{\mb I}}

\newcommand{\1}{1 \!\! 1}

\newcommand{\Hil}{\mc H}

\newtheorem{thm}{Theorem}

\newtheorem{prop}[thm]{Proposition}

\catcode `\@=11 \@addtoreset{equation}{section}
\def\theequation{\arabic{section}.\arabic{equation}}
\catcode `\@=12

\begin{document}

\begin{center}
{\Large \bf Stock markets and quantum dynamics: a second quantized description}   \vspace{2cm}\\

{\large F. Bagarello}
%\footnote[1]{ Dipartimento di Matematica ed Applicazioni,
%Fac.Ingegneria, Universit\`a di Palermo, I - 90128  Palermo, Italy}
\vspace{3mm}\\
  Dipartimento di Metodi e Modelli Matematici,
Facolt\`a di Ingegneria,\\ Universit\`a di Palermo, I - 90128  Palermo, Italy\\
E-mail: bagarell@unipa.it\\home page:
www.unipa.it$\backslash$\~\,bagarell
\vspace{2mm}\\
\end{center}

\vspace*{2cm}

\begin{abstract}
\noindent In this paper we continue our descriptions of stock
markets in terms of some non abelian operators which are used to
describe the portfolio of the various traders and other {\em
observable} quantities. After a first prototype model with only two
traders, we discuss a more realistic model of market with an
arbitrary number of traders. For both models we find approximated
solutions for the time evolution of the portfolio of each trader. In
particular, for the more realistic model, we use the {\em stochastic
limit} approach and a {\em fixed point like} approximation.

\end{abstract}

\vfill

\newpage

% Section 1
\section{Introduction}

In a recent paper, \cite{bag1}, we have discussed how a quantum
mechanical framework can be used in the analysis of stock markets.
The conservation of the total number of shares and of the total
amount of cash in any {\bf closed} marked, i.e. in a market which
does not {\em interact} with the environment, as well as the
discrete nature of the number of the shares and of the {\em monetary
unit}, suggested the use of some typical tools of $QM_\infty$, i.e.
of quantum mechanics for systems with infinite degrees of freedom,
in this different context. In particular, we have shown that a {\em
second quantized vision} of the stock market produces in a natural
way a set of differential equations describing the time evolution of
the portfolio of each trader of the market. These results are on the
same line as those given in \cite{mar} and \cite{scha1}, as well as
in \cite{baa} and references therein. We should also mention that
the use of tools coming from physics for economical problems, or
more generally for dealing with {\em complex systems}, is a well
established procedure, for which we refer to \cite{sasa}.

The paper is organized as follows:

in the next section we briefly review the results in \cite{bag1}.

In Section III, contrarily with what has been made in \cite{bag1},
where the price of the share $P(t)$ has no {\em real} dynamics since
it is replaced by its mean value, we introduce a prototype model
with only two traders were we show how to keep into account the time
evolution of the price $P(t)$. We prove that many integrals of
motion exist. The equations of motion are solved using a
perturbative expansion, well known in $QM_\infty$.

In Section IV we consider another model, which generalize the
previous one in the sense that it consists of $N$ traders with
arbitrary $N\geq 2$, and for which we consider two different
approximations: the stochastic limit, which is useful to analyze the
equilibrium of the model, and what we call a {\em fixed point-lixe}
(FPL) approximation, which we use to deduce the approximated time
evolution of the portfolio of any given trader of the stock market
associated to the model.

Section V contains our conclusions and plans for the future, while
we discuss in the first Appendix some delicate mathematical points
and in Appendix B a general introduction to the stochastic limit.

\section{The genesis of the model}

In this section we review some results and ideas first introduced in
\cite{bag1} which have produced an interesting toy model of a stock
market based on the following assumptions:
\begin{enumerate}
\item Our market consists of $L$ traders exchanging a single kind
of share; \item the total number of shares, $N$, is fixed in time;
\item a trader can only interact with a single other trader: i.e.
the traders feel only a {\em two-body interaction};\item the traders
can only buy or sell one share in any single transaction;\item the
price of the share changes with discrete steps, multiples of a given
monetary unit; \item when the  tendency of the market to sell a
share, i.e. the {\em market supply}, increases then the price of the
share decreases; \item for our convenience the  supply is expressed
in term of natural numbers; \item to simplify the notation, we take
the monetary unit equal to 1.
\end{enumerate}

We refer to \cite{bag1} for the analysis of these conditions, which
however, in our opinion, look quite natural and self-explanating.
The {\em formal} hamiltonian of the model is the following operator:
\be \left\{
\begin{array}{ll}
\tilde H=H_0+\tilde H_I, \mbox{ where }  \\
H_0 = \sum_{l=1}^L\alpha_l a_l^\dagger a_l+\sum_{l=1}^L\beta_l
c_l^\dagger c_l+o^\dagger \,o+p^\dagger \,p \\
\tilde H_I = \sum_{i,j=1}^Lp_{ij}\left( a_i^\dagger a_j
(c_i\,c_j^\dagger)^{\hat P}+\,a_i\, a_j^\dagger
(c_j\,c_i^\dagger)^{\hat P}\right)+(o^\dagger \,p+p^\dagger \,o), \\
\end{array}
\right. \label{31} \en where $\hat P=p^\dagger p$ and the following
commutation rules are assumed: \be
[a_l,a_n^\dagger]=[c_l,c_n^\dagger]=\delta_{ln}\id, \hspace{.5cm}
[p,p^\dagger]=[o,o^\dagger]=\id,\label{32}\en while all the other
commutators are zero. We further assume that $p_{ii}=0$. Here the
operators $a_l^\sharp$, $p^\sharp$, $c_l^\sharp$ and $o^\sharp$ are
respectively the {\em number}, the {\em price}, the {\em cash} and
the {\em supply} operators, \cite{bag1}. The {\em states} of the
market are \be
\omega_{\{n\};\{k\};O;M}(\,.\,)=<\varphi_{\{n\};\{k\};O;M},
\,.\,\varphi_{\{n\};\{k\};O;M}>,\label{33}\en where
$\{n\}=n_1,n_2,\ldots,n_L$, $\{k\}=k_1,k_2,\ldots,k_L$ and \be
\varphi_{\{n\};\{k\};O;M}:=\frac{(a_1^\dagger)^{n_1}\cdots
(a_L^\dagger)^{n_L}(c_1^\dagger)^{k_1}\cdots
(c_L^\dagger)^{k_L}(o^\dagger)^O(p^\dagger)^M}{\sqrt{n_1!\ldots
n_L!\,k_1!\ldots k_L!\,O!\,M!}}\,\varphi_0. \label{34}\en Here
$\varphi_0$ is the {\em vacuum } of the model:
$a_j\varphi_0=c_j\varphi_0=p\varphi_0=o\varphi_0=0$, for
$j=1,2,\ldots,L$. Again we refer to \cite{bag1} or to any quantum
mechanical textbook, see \cite{mer} for instance, for further
details on second quantization.

The interpretation of the hamiltonian is a key point in our approach
and has been discussed in details in \cite{bag1}: just as an
example, the presence of the term $o^\dagger \,p$ in $\tilde H_I$
implies that when the supply increases then the price must decrease.
Moreover, because of  $a_i^\dagger a_j (c_i\,c_j^\dagger)^{\hat P}$,
 trader $\tau_i$ increases of one unit the number of shares in his
portfolio but, at the same time, his cash decreases because of
$c_i^{\hat P}$, that is it must decrease of as many units of cash as
the price operator $\hat P$ demands. Clearly,  trader $\tau_j$
behaves in the opposite way: he loses one  share because of $a_j$
but his cash increases because of $(c_j^\dagger)^{\hat P}$.

However we have discussed in \cite{bag1} that the hamiltonian in
(\ref{31}) suffers of a technical problem: since $c_j$ and
$c_j^\dagger$ are not self-adjoint operators, it was not obvious how
to define the operators $c_j^{\hat P}$ and $(c_j^\dagger)^{\hat P}$,
and for this reason we have replaced $\tilde H$ with an {\em
effective } hamiltonian, $H$, defined as

\be \left\{
\begin{array}{ll}
 H=H_0+H_I, \mbox{ where }  \\
 H_0 = \sum_{l=1}^L\alpha_l a_l^\dagger a_l+\sum_{l=1}^L\beta_l
c_l^\dagger c_l+o^\dagger \,o+p^\dagger \,p \\
H_I = \sum_{i,j=1}^Lp_{ij}\left( a_i^\dagger a_j
(c_i\,c_j^\dagger)^{M}+\,a_i\, a_j^\dagger
(c_j\,c_i^\dagger)^{M}\right)+(o^\dagger \,p+p^\dagger \,o), \\
\end{array}
\right. \label{35} \en where $M=\omega_{\{n\};\{k\};O;M}(\hat P)$.
In this way, however, we are essentially {\em freezing} the price of
our action, removing one of the (essential) degrees of freedom out
from our market. This strong limitation will be removed in the next
two sections of this paper and, in our opinion, this is really a
major improvement.

 Three integrals of motion for our model
trivially exist: \be \hat N=\sum_{i=1}^L
a_i^\dagger\,a_i,\hspace{3mm}\hat K=\sum_{i=1}^Lc_i^\dagger
c_i\hspace{3mm}\mbox{ and }\hspace{3mm} \hat\Gamma =o^\dagger
o+p^\dagger p.\label{36}\en This can be easily checked since the
canonical commutation relations in (\ref{32}) imply that $[H,\hat
N]=[H,\hat\Gamma]= [H,\hat K]=0$.

The fact that $\hat N$ is conserved clearly means that no new share
is introduced in the market. Of course, also the total amount of
money must be a constant of motion since  the cash is assumed to be
used only to buy shares. Since also $\hat \Gamma$ commutes with $H$,
moreover, if the mean value of $o^\dagger o$ increases with time
then necessarily the mean value of the price operator $\hat
P=p^\dagger p$ must decrease and vice-versa. This is exactly the
mechanism assumed in point { 6.} at the beginning of this section.
Moreover, also the following operators commute with $H$ and, as a
consequence, are constant in time: \be \hat Q_j=a_j^\dagger
\,a_j+\frac{1}{M}\,c_j^\dagger \,c_j,\label{37}\en for
$j=1,2,\ldots,L$.

The hamiltonian (\ref{35}) contains a contribution,
$h_{po}=o^\dagger \,o+p^\dagger \,p+(o^\dagger \,p+p^\dagger \,o)$,
which is decoupled from the other terms. For this reason it is easy
to deduce the time dependence of both the price and the supply
operators, as well as  of their mean values. We get, \cite{bag1},
 \be \left\{
\begin{array}{ll}
P_r(t) = \frac{1}{2}\{P_r+O+(P_r-O)\cos(2t)\} \\
O(t) = \frac{1}{2}\{P_r+O-(P_r-O)\cos(2t)\}, \\
\end{array}
\right. \label{37b} \en where we have called
$O(t)=\omega_{\{n\};\{k\};O;M}(o^\dagger(t) o(t))$ and
$P_r(t)=\omega_{\{n\};\{k\};O;M}(p^\dagger(t) p(t))$. Recall that
$P_r=P_r(0)=M$. Equations (\ref{37b}) show that, if $O=P_r$ then
$O(t)=P_r(t)=O$ for all $t$ while, if $O\simeq P_r$ then $O(t)$ and
$P_r(t)$ are {\em almost } constant. In the following we will
replace $P_r(t)$ with an integer value, the value $M$ which appears
in the hamiltonian (\ref{35}), which is therefore fixed
\underline{after} the solution (\ref{37b}) is found. This value is
obtained by taking a suitable mean of $P_r(t)$ or working in one of
the following assumptions: (i) $O=P_r$; or (ii) $O\simeq P_r$ or yet
(iii) $|O+P_r|\gg|P_r-O|$. In these last two situations we may
replace $P_r(t)$, with a temporal mean, $<P_r(t)>$, since there is
not much difference between these two quantities. However, in this
way we are essentially removing the dynamics of the price from the
model: no price variation occurs within this model after the
replacement $P_r(t)\rightarrow M$! How already anticipated, this
restriction will be removed in the next section.

Our main result in \cite{bag1} was to deduct the time evolution of
the {\em portfolio} operator, which we have defined as  \be \hat
\Pi_j(t)=\gamma\hat n_j(t)+\hat k_j(t).\label{38}
 \en
Here we have introduced the value of the share $\gamma$ {\em as
decided by the market}, which does not necessarily coincides with
the amount of money which is payed to buy the share. As it is clear,
$\hat \Pi_j(t)$ is the sum of the complete value of the shares, plus
the cash. The fact that for each $j$ the operator $Q_j$ is an
integral of motion allows us to rewrite the operator $\hat \Pi_j(t)$
only in terms of $\hat n_j(t)$ and of the initial conditions. We
find: \be \hat \Pi_j(t)=\hat\Pi_j(0)+(\gamma-M)(\hat n_j(t)-\hat
n_j(0)),\label{39}
 \en
In order to get the time behavior of the portfolio, therefore, it is
enough to obtain $\hat n_j(t)$. We refer to \cite{bag1} for a simple
perturbative expansion of $\Pi_j(t)$ for $L=2$. Here we prefer to
show the other results, also contained in \cite{bag1}, concerning
the {\em semiclassical thermodynamical limit} of the model, i.e. a
suitable limit for $L\rightarrow\infty$.

Our model is defined by the same hamiltonian as in (\ref{35}) but
with $M=1$. This is not a major requirement here since it
corresponds to a renormalization of the price of the share, which we
take equal to 1: if you buy a share, then your liquidity decreases
of one unit while it increases, again of one unit, if you sell a
share. Needless to say, this is strongly related to the fact that
the original time-dependent price operator $\hat P(t)$ has been
replaced by a certain weak mean value, $M$.

It is clear that all the same integrals of motion as before
 exist: $\hat N$, $\hat K$, $\hat\Gamma$, $\hat\Delta:=o-p$ and
$Q_j=\hat n_j+\hat k_j$, $j=1,2,\ldots,L$. They all commute with
$H$, which we now write as \be \left\{
\begin{array}{ll}
 H=h+h_{po}, \mbox{ where }  \\
 h = \sum_{l=1}^L\alpha_l \hat n_l+\sum_{l=1}^L\beta_l
\hat k_l +\sum_{i,j=1}^Lp_{ij}\left( a_i^\dagger a_j
c_i\,c_j^\dagger+\,a_i\, a_j^\dagger
c_j\,c_i^\dagger\right)\\
h_{po} = o^\dagger \,o+p^\dagger \,p+(o^\dagger \,p+p^\dagger \,o), \\
\end{array}
\right. \label{41} \en For $h_{po}$ we can repeat the same argument
as above and an explicit solution can be found which is completely
independent of $h$. In particular we have
$\omega_{\{n\};\{k\};O;M}(\hat P)=1$.
 For
this reason, from now on, we will identify $H$ only with $h$ in
(\ref{41}) and we will work only with this hamiltonian. Let us
introduce the operators\be X_i=a_i\,c_i^\dagger,\label{42}\en
$i=1,2,\ldots,L$. The hamiltonian $h$ can be rewritten as \be
h=\sum_{l=1}^L\left(\alpha_l \hat n_l+\beta_l \hat k_l\right)
+\sum_{i,j=1}^L
p_{ij}\left(X_i^\dagger\,X_j+X_j^\dagger\,X_i\right).\label{43}\en
The following commutation relations hold: \be
 [X_i,X_j^\dagger]=\delta_{ij}(\hat k_i-\hat n_i), \hspace{5mm}  [X_i,\hat n_j]=\delta_{ij}\,X_i
 \hspace{5mm}  [X_i,\hat k_j]=-\delta_{ij}\,X_i,  \label{44} \en
which show how the operators $\{\{X_i,\,X_i^\dagger,\,\hat
n_i,\,\hat k_i\},\,i=1,2,\ldots,L\}$ are closed under commutation
relations. This is quite  important, since, introducing the
operators $X_l^{(L)}=\sum_{i=1}^Lp_{li}X_i$, $l=1,2,\ldots,L$, we
get the following system of  differential equations, see
\cite{bag1}:
 \be \left\{
\begin{array}{ll}
 \dot X_l=i(\beta_l-\alpha_l)X_l+2iX_l^{(L)}(2\hat n_l-Q_l)  \\
\dot{\hat n_l}=2i\left(X_l\,{X_l^{(L)}}^\dagger-X_l^{(L)}\,X_l^\dagger\right)\\
\end{array}
\right. \label{45} \en This system, as $l$ takes all the values
$1,2,\ldots,L$, is a closed system of differential equations for
which an unique solution surely exists. Indeed, we have found such a
solution in \cite{bag1} by introducing the so-called mean-field
approximation which essentially consists in replacing  $p_{ij}$ with
$\frac{\tilde p}{L}$, with $\tilde p\geq 0$. After this replacement
we have that
$$X_l^{(L)}=\sum_{i=1}^Lp_{li}X_i\longrightarrow \frac{\tilde
p}{L}\sum_{i=1}^LX_i, $$ whose limit, for $L$ diverging, only exists
in suitable topologies, \cite{thi,bm}, like, for instance, the
strong one restricted to a set of relevant states. Let $\tau$ be
such a topology. We define\be
X^{\infty}=\tau-\lim_{L\rightarrow\infty}\frac{\tilde
p}{L}\sum_{i=1}^LX_i,\label{46}\en where, as it is clear, the
dependence on the index $l$ is lost because of the replacement
$p_{li}\rightarrow\frac{\tilde p}{L}$. This is a typical behavior of
transactionally invariant quantum systems, where $p_{l,i}=p_{l-i}$.
The operator $X^\infty$  commutes with all the elements of $\A$, the
{\em algebra of the observables} of our stock market:
$[X^\infty,A]=0$ for all $A\in\A$. In this limit system (\ref{45})
above becomes \be \left\{
\begin{array}{ll}
 \dot X_l=i(\beta_l-\alpha_l)X_l+2iX^{\infty}(2\hat n_l-Q_l)  \\
\dot{\hat n_l}=2i\left(X_l\,{X^{\infty}}^\dagger-X^{\infty}\,X_l^\dagger\right)\\
\end{array}
\right. \label{47} \en  This system has been solved in \cite{bag1}
under the hypothesis that \be\beta_l-\alpha_l=:\Phi\neq
\nu\label{48}\en for all $l=1,2,\ldots,L$ (but also in other and
more general situations). We again refer to \cite{bag1} for the
details. Here we just write the final result, which is \be
n_l(t)=\frac{1}{\omega^2}\left\{n_l(\Phi-\nu)^2-8|X_0^\infty|^2\left(k_l(\cos(\omega
t)-1)-n_l(\cos(\omega t)+1)\right)\right\},\label{414}\en where we
have introduced $\omega=\sqrt{(\Phi-\nu)^2+16|X_0^\infty|^2}$. This
allows also to find the time evolution for the portfolio, since
$\Pi_l(t)=\Pi_l(0)+(\gamma-1)(n_l(t)-n_l(0))$. Again, we refer to
\cite{bag1} for further comments and results. Here we just want to
stress that our point of view has really produced, as an output, the
time evolution of the portfolio of each trader of the market, which
was indeed our original aim.

\section{A two trader model}

As we have already discussed the model analyzed in the previous
section has  a very strong limitation: the time evolution of the
price of the share, even if formally appears in the hamiltonian of
the system, is frozen in order to get a well defined {\em energy
operator} (i.e. in moving from $\tilde H$ to $H$). Therefore, and in
particular when we consider the thermodynamical limit of the model,
such a dynamical behavior of the price operator completely
disappears!

In this section we will cure this anomaly, and for that we will
discuss in many details a model based essentially on the same
assumptions listed at the begin of   Section II but in which the
market consists of only two traders, $\tau_1$ and $\tau_2$. Of
course, more than a realistic stock market, this can be seen as a
sort of two-components physical system ($\tau_1+\tau_2$) changing
two different kind of particles (the shares and the money) and
subjected to an external control (the price of the share and the
supply of the system itself). However, even in view of the
generalization which we will discuss in the next section, we will
still refer to this physical system as a (toy model of a) stock
market.

The hamiltonian looks very much as the one in (\ref{31}): \be
\left\{
\begin{array}{ll}
H=H_0+ H_I, \mbox{ where }  \\
H_0 = \sum_{l=1}^2\alpha_l a_l^\dagger a_l+\sum_{l=1}^L\beta_l
c_l^\dagger c_l+o^\dagger \,o+p^\dagger \,p \\
\tilde H_I = \left( a_1^\dagger a_2 c_1^{\hat
P}\,{c_2^\dagger}^{\hat P}+\,a_1\, a_2^\dagger
{c_1^\dagger}^{\hat P}\,c_2^{\hat P}\right)+(o^\dagger \,p+p^\dagger \,o), \\
\end{array}
\right. \label{51} \en with the standard commutation relations \be
[o,o^\dagger]=[p,p^\dagger]=\1,\qquad
[a_i,a_j^\dagger]=[a_i,a_j^\dagger]=\delta_{i,j}\1,\label{52}\en
while all the other commutators are zero.

The states of the system are defined as in (\ref{33}) and (\ref{34})
with $L=2$, and the vectors $\varphi_{\{n\};\{k\};O;M}$ are
eigenstates of the operators $\hat n_i=a_i^\dagger a_i$, $\hat
k_i=c_i^\dagger c_i$, $i=1,2$, $\hat P=p^\dagger p$ and $\hat
\Omega=o^\dagger o$, respectively with eigenvalues $n_i, k_i$,
$i=1,2$, $M$ and $O$. The main achievement here is that, how we
discuss in Appendix A, we are now able to give a rigorous meaning to
the operators $c_j^P$ and ${c_j^\dagger}^P$, and this  allow us {\bf
not} to replace the price operator with its mean value $M$ and, as a
consequence, to consider the price of the share as a real {\em
degree of freedom} of the model. However, before defining $c_j^P$
and ${c_j^\dagger}^P$, it is worth noticing that the non abelianity
of our structure does not automatically implies that the observables
of the market, i.e. the operators $\hat k_i$, $\hat n_i$, $\hat P$
and $\hat\Omega$, as well as some of their combinations, cannot be
measured simultaneously. This is because these observables, which
are the only relevant variables for us,  do commute and, as a
consequence, they admit a common set of eigenstates, see equation
(\ref{53}) below.

Using the same arguments given in  Appendix A we are able to define
the operators $c_j^P$ and ${c_j^\dagger}^P$ via their action on the
orthonormal (o.n.) basis of the Fock-Hilbert space $\Hil$ of the
model whose generic vector is, in analogy with (\ref{34}), \be
\varphi_{n_1,n_2;\,k_1,k_2;\,O;\,M}:=\frac{(a_1^\dagger)^{n_1}
(a_2^\dagger)^{n_2}(c_1^\dagger)^{k_1}
(c_2^\dagger)^{k_2}(o^\dagger)^O(p^\dagger)^M}{\sqrt{n_1!\,
n_2!\,k_1!\, k_2!\,O!\,M!}}\,\varphi_0. \label{53}\en Here $n_j$,
$k_j$, $O$ and $M$ are non negative integers, $\varphi_0$ is the
{\em vacuum } of the model,
$a_j\varphi_0=c_j\varphi_0=p\varphi_0=o\varphi_0=0$, for $j=1,2$,
and $\Hil$ is the closure of the linear span of all these vectors.
Then we have, for instance,
$a_1\varphi_{n_1,n_2;\,k_1,k_2;\,O;\,M}=\sqrt{n_1}\,\varphi_{n_1-1,n_2;\,k_1,k_2;\,O;\,M}$
if $n_1>0$ and $a_1\varphi_{n_1,n_2;\,k_1,k_2;\,O;\,M}=0$ if
$n_1=0$,
$a_1^\dagger\varphi_{n_1,n_2;\,k_1,k_2;\,O;\,M}=\sqrt{n_1+1}\,\varphi_{n_1+1,n_2;\,k_1,k_2;\,O;\,M}$.
Analogous expressions for the action of $a_2$, $a_2^\dagger$, $c_j$,
$c_j^\dagger$, $o$, $o^\dagger$, $p$ and $p^\dagger$ on
$\varphi_{n_1,n_2;\,k_1,k_2;\,O;\,M}$ can also be recovered, see
\cite{mer}. Moreover we have, see Appendix A, \bea c_1^{\,\hat
P}\,\varphi_{n_1,n_2;\,k_1,k_2;\,O;\,M}:= \left\{
\begin{array}{ll}
\varphi_{n_1,n_2;\,k_1,k_2;\,O;\,M},  &\mbox{ if } M=0,\,\forall k_1\geq 0;  \\
0,  &\mbox{ if } M>k_1,\,\forall k_1\geq 0;\\
\sqrt{k_1^{\{-M\}}}\,\varphi_{n_1,n_2;\,k_1-M,k_2;\,O;\,M}, & \mbox{
if } k_1\geq M
> 0
\end{array}
\right. \label{54}\ena and \bea {c_1^\dagger}^{\,\hat
P}\,\varphi_{n_1,n_2;\,k_1,k_2;\,O;\,M}:=\left\{
\begin{array}{ll}
\varphi_{n_1,n_2;\,k_1,k_2;\,O;\,M},  &\mbox{ if } M=0,\,\forall k_1\geq 0;  \\
\sqrt{k_1^{\{+M\}}}\,\varphi_{n_1,n_2;\,k_1+M,k_2;\,O;\,M}, & \mbox{
if } M>0,
\end{array}
\right. \label{55}\ena where we have defined \be\left\{
\begin{array}{ll}
k_1^{\{-M\}}:=k_1(k_1-1)\cdots(k_1-M+1)\\
k_1^{\{+M\}}:=(k_1+1)(k_1+2)\cdots(k_1+M)
\end{array}
\right. \label{56}\en Analogous formulas hold for $c_2^{\,\hat P}$
and ${c_2^\dagger}^{\,\hat P}$. These definitions have a clear {\em
economical} interpretation: acting with $c_1^{\,\hat P}$ on
$\varphi_{n_1,n_2;\,k_1,k_2;\,O;\,M}$ returns
$\varphi_{n_1,n_2;\,k_1,k_2;\,O;\,M}$ itself when $M=0$ since, in
this case, the action of $c_1^{\,\hat P}$ coincides with that of the
identity operator: the price of the share is zero so you don't need
to pay for it and hence your cash does not change! Moreover, if
$M>k_1$, $c_1^{\,\hat P}$ destroys more {\em quanta of money} than
$\tau_1$ really possesses. Therefore, the result of its action on
the vector is zero. A similar problem does not occur when we
consider the action of ${c_1^\dagger}^{\,\hat P}$ on
$\varphi_{n_1,n_2;\,k_1,k_2;\,O;\,M}$, since in this case the cash
is created!

In the rest of the section, however, these formulas will be
significantly simplified by assuming that, as it is reasonable,
during the transactions between $\tau_1$ and $\tau_2$ the price of
the share never reach the zero value and, moreover, that no trader
try to buy a share if he has not enough money to pay for it.
Therefore we simply rewrite (\ref{54}) and (\ref{55}) as \be\left\{
\begin{array}{ll}
c_1^{\,\hat P}\,\varphi_{n_1,n_2;\,k_1,k_2;\,O;\,M}=
\sqrt{k_1^{\{-M\}}}\,\varphi_{n_1,n_2;\,k_1-M,k_2;\,O;\,M},\\
{c_1^\dagger}^{\,\hat
P}\,\varphi_{n_1,n_2;\,k_1,k_2;\,O;\,M}=\sqrt{k_1^{\{+M\}}}\,
\varphi_{n_1,n_2;\,k_1+M,k_2;\,O;\,M}\\
\end{array}
\right. \label{57}\en The commutation rules are the standard ones,
see (\ref{32}), plus the ones which extend the rules in
(\ref{app4}): \be [{\hat P},c_j^{\,\hat P}]=[{\hat
P},{c_j^\dagger}^{\,\hat P}]=0, \label{58} \en and \be [\hat
k_j,c_l^{\,\hat P}]=-\delta_{j,l}\,{\hat P}\,c_j^{\,\hat P},\qquad
[\hat k_j,{c_l^\dagger}^{\,\hat P}]=\delta_{j,l}\,{\hat
P}\,{c_j^\dagger}^{\,\hat P}\label{59} \en  for $j=1,2$.

Since our market is closed it is not surprising that the total
number of shares and the total amount of cash are preserved. This is
indeed proved simply computing the commutators of the total number
of shares and the total cash operators, $\hat N=\hat n_1+\hat n_2$
and $\hat K=\hat k_1+\hat k_2$, with the hamiltonian $H$. Indeed one
can check that $[H,\hat K]=[H,\hat N]=0$. Moreover, we can also
check that $\hat \Gamma:=\hat \Omega+\hat P$ commutes with the
hamiltonian. This is, as already discussed in the previous section,
the mechanism which fixes the price of the share within our
simplified market: the more the market supply increases the less is
the value of the share, i.e. its price.

As already stressed before, one big difference between the model we
are considering here and the one considered in Section II and in
\cite{bag1} is that now the price operator is not replaced by its
mean value. This has an important consequence: the operators
extending $Q_j$ in (\ref{37}) for this model, which are proportional
to $\hat P\,a_j^\dagger \,a_j+\,c_j^\dagger \,c_j$, $j=1,2$, are no
longer constants of motion, and they cannot be used to facilitate
the computation of the portfolios of $\tau_1$ and $\tau_2$.
Nevertheless we will still be able to deduce, with an easy
perturbative approach, the time behavior of both portfolios at least
for small values of $t$.

The first step consists in deducing the time evolution of the price
of the share. This computation is completely analogous to that of
\cite{bag1} and will not be repeated here. Again, we can deduce that
$\hat \Delta:=o-p$ is another constant of motion and we find that,
see (\ref{37b}),
 \be \left\{
\begin{array}{ll}
P(t) =\omega_{n_1,n_2;\,k_1,k_2;\,O;\,M}(p^\dagger(t) p(t))=\frac{1}{2}\{M+O+(M-O)\cos(2t)\} \\
O(t) =\omega_{n_1,n_2;\,k_1,k_2;\,O;\,M}(o^\dagger(t) o(t))= \frac{1}{2}\{M+O-(M-O)\cos(2t)\} \\
\end{array}
\right. \label{510} \en In \cite{bag1} the absence of a true
dynamics for $\hat P$ suggested to define the portfolio of the
$j-th$ trader by introducing another parameter, $\gamma$, which was
interpreted as the price of the share {\em as decided by the
market}, which does not necessarily coincides with $M$. However,
there was no direct link between $\gamma$ and $M$ in \cite{bag1},
and this is not completely satisfying, of course! Here we have no
need for introducing such an extra parameter since we are now in a
position to consider directly $P(t)$ instead of its mean value $M$.
Therefore we replace formula (\ref{38}) by defining the portfolio of
the trader $\tau_j$ as \be \hat \Pi_j(t)=\hat P(t)\,\hat n_j(t)+\hat
k_j(t)\label{511}\en with $j=1,2$, which is just the sum of the
total price of the shares and the cash of $\tau_j$.

Of course, due to the fact that $P(t)$ is known, $\hat \Pi_1(t)$ is
known when we both know $\hat n_1(t)$ and $\hat k_1(t)$. Moreover,
if we now $\hat n_1(t)$ and $\hat k_1(t)$, then we also know $\hat
n_2(t)$ and $\hat k_2(t)$ since their sum must be constant, so that
we can also find the analytic form of $\hat\Pi_2(t)$. However, this
is not the only way to find $\hat \Pi_1(t)$. Another possibility
follows from the fact that, as it is easy to check, \be \dot {\hat
\Pi}_1(t)=\dot{\hat P}(t)\,\hat n_1(t),\label{512}\en which shows
again, even without any need of using $Q_j$ as in the previous
section, that it is enough to know $\hat n_1(t)$ to find the time
evolution of the portfolio of $\tau_1$.

However, even for this two-traders model, it is not easy to deduce
the exact expression for $\hat \Pi_1(t)$. Nevertheless,  a lot of
information can be deduced, mainly for short time behavior, using
different perturbative strategies. Here we just consider the most
{\em direct} technique, i.e. the following perturbative expansion
\be \hat \Pi_1(t)=e^{iHt}\hat \Pi_1(0)e^{-iHt}=\hat
\Pi_1(0)+it[H,\hat \Pi_1(0)]+\frac{(it)^2}{2!}[H,[H,\hat
\Pi_1(0)]]+\cdots, \label{512bis}\en leaving to the next section a
more detailed analysis of other strategies to produce
$\hat\Pi_j(t)$.
 The computation of the various
terms of this expansion, and of their mean values on the state
$\omega_{n_1,n_2;\,k_1,k_2;\,O;\,M}(.)$, is based on the commutation
rules we have seen before and produce, up to the second order in
$t$, the following result:
$$
\Pi_1(t)=\omega_{n_1,n_2;\,k_1,k_2;\,O;\,M}(\hat
\Pi_1(t))=\Pi_1(0)+t^2n_1(O-M),
$$
which shows that, for sufficiently small values of $t$, the value of
$\Pi_1(t)$ increases with time if $O>M$, i.e. if at $t=0$ and in our
units the supply of the market is larger than the price of the
share. It is further possible to check that the next term in the
expansion above is proportional to $t^4\,A_{n_1,n_2;k_1,k_2;M}$
where $$A_{n_1,n_2;k_1,k_2;M}=
n_1\,k_1^{(+M)}\,k_2^{(-M)}-n_2\,k_1^{(-M)}\,k_2^{(+M)}+n_1n_2(k_1^{(+M)}\,k_2^{(-M)}-
k_1^{(-M)}\,k_2^{(+M)}).$$ We avoid the details of this computation
here since they are not very interesting, mainly because this is
just a toy model which is more important for its general structure
than for a real financial interpretation. Here we just want to
stress that the expansion in (\ref{512bis}) gives, in principle, the
expression of $\hat \Pi_1(t)$ at any desired approximation.

\section{Many traders}

In the previous section we have learned how to define  the operators
$c^{\,\hat P}$ and its adjoint and we have used this definition in
the analysis of a simple hamiltonian which was essentially already
introduced in \cite{bag1}. We devote this section to a more
realistic model, where the stock market is made of $N$ different
traders with $N$ arbitrarily large.

In our approach we will focus  our attention on a single trader,
$\tau$, which interact with an ensemble of other traders in  a way
that extends the interaction introduced in (\ref{51}). In other
words we divide the stock market, which as before is defined in
terms of the number of a single type of shares, the cash, the price
of the shares and the supply of the market, in two main {\em
ingredients}: we call  {\em system}, $\Sc$, all the dynamical
quantities which refer to a fixed trader $\tau$: its {\em shares
number operators}, $a$, $a^\dagger$ and $\hat n=a^\dagger\,a$, the
{\em cash operators} of $\tau$, $c$, $c^\dagger$ and $\hat
k=c^\dagger\,c$ as well as the {\em price operators} of the shares,
$p$, $p^\dagger$ and $\hat P=p^\dagger\,p$. On the other hand, we
associate to the {\em reservoir}, $\R$, all the other quantities,
that is first of all, the {\em shares number operators}, $A_k$,
$A_k^\dagger$ and $\hat N_k=A_k^\dagger\,A_k$ and the {\em cash
operators}, $C_k$, $C_k^\dagger$ and $\hat K_k=C_k^\dagger\,C_k$ of
the other traders. Here $k\in\Lambda$ and $\Lambda$ is a subset of
$\Bbb{N}$ which labels the traders of the market (other than
$\tau$). It is clear that the cardinality of $\Lambda$ is $N-1$.
Moreover we associate to the reservoir also the supply of the
market, which is described by the following operators $o_k$,
$o_k^\dagger$ and $\hat O_k=o_k^\dagger\,o_k$, $k\in\Lambda$. The
stock market is given by the union of $\Sc$ and $\R$, and the
hamiltonian, which extends the one in (\ref{51}), is assumed here to
be
 \be
\left\{
\begin{array}{ll}
H=H_0+ \lambda\,H_I, \mbox{ where }  \\
H_0 = \omega_a\, \hat n+\omega_c\, \hat k+\omega_p \hat
P+\sum_{k\in\Lambda}\left(\Omega_A(k)\,\hat N_k+
 \Omega_C(k)\,\hat K_k+\Omega_O(k)\,\hat  O_k\right)\\
 H_I = \left( z^\dagger \,Z(f)+z\,Z^\dagger(\overline f)\right)+(p^\dagger \,o(g)+p \,o^\dagger(\overline g)) \\
\end{array}
\right. \label{61} \en Here $\omega_a$, $\omega_c$ and $\omega_p$
are positive real numbers and $\Omega_A(k)$, $\Omega_C(k)$ and
$\Omega_O(k)$ are real valued non negative functions, whose
interpretation was first discussed in \cite{bag1}: they describe the
free time evolution of the different operators of the market. We
have also introduced the following {\em smeared fields} of the
reservoir:
\be\left\{\begin{array}{ll}Z(f)=\sum_{k\in\Lambda}Z_k\,f(k)=\sum_{k\in\Lambda}A_k\,{C_k^\dagger}^{\,\hat
P}\,f(k), \\ Z^\dagger(\overline
f)=\sum_{k\in\Lambda}Z_k^\dagger\,\overline{f(k)}=\sum_{k\in\Lambda}A_k^\dagger\,{C_k}^{\,\hat
P}\,\overline{f(k)}\\
o(g)=\sum_{k\in\Lambda}o_k\,g(k) \\ o^\dagger(\overline
g)=\sum_{k\in\Lambda}o_k^\dagger\,\overline{g(k)},\end{array}\right.\label{61bis}\en
as well as the operators $z=a\,{c^\dagger}^{\,\hat P}$,
$Z_k=A_k\,{C_k^\dagger}^{\,\hat P}$ and their conjugates, since for
instance $A_k$ and $C_k$ appear always in this combination both in
$H_I$ and in all the computations we will perform in the following.
This is natural because of the {\em physical} meaning of, e.g., $z$:
the action of $z$ on a fixed vector number destroys a share in the
portfolio of $\tau$ and, at the same time, creates as many monetary
units as $\hat P$ prescribes! Of course, in $H_I$ such an operator
is associated to $Z^\dagger(\overline f)$ which acts exactly in the
opposite way on the traders of the reservoir: one share is created
in the cumulative portfolio of $\R$ while $\hat P$ {\em quanta} of
money are destroyed, since they are used to pay for the share. The
following non trivial commutation rules are assumed: \be
[c,c^\dagger]=[p,p^\dagger]=[a,a^\dagger]=\1,\qquad
[o_i,o_j^\dagger]=[A_i,A_j^\dagger]=[C_i,C_j^\dagger]=\delta_{i,j}\1\label{62}\en
which implies \be [\hat K_k,C_q^{\,\hat P}]=-\hat P\,C_q^{\,\hat
P}\,\delta_{k,q},\quad  [\hat K_k,{C_q^\dagger}^{\,\hat P}]=\hat
P\,{C_q^{\dagger}}^{\,\hat P}\,\delta_{k,q} \label{63}\en

Finally, the functions $f(k)$ and $g(k)$ in (\ref{61}) and
(\ref{61bis}) are sufficiently regular to allows for the  sums in
(\ref{61bis}) to be well defined, as well as the quantities which
will be defined below, see (\ref{69}).

\vspace{2mm} {\bf Remark:--} Of course, since $\tau$ can be chosen
arbitrarily, the asymmetry of the model is just apparent. In fact,
changing $\tau$, we will be able, in principle, to find the time
evolution of the portfolio of each trader of the stock market.

\vspace{2mm}

The interpretation suggested above concerning $z$ and $Z(f)$ are
also based on the following results: let \be \hat N:=\hat
n+\sum_{k\in\Lambda}\,\hat N_k,\quad \hat K:=\hat
k+\sum_{k\in\Lambda}\,\hat K_k,\quad \hat \Gamma:=\hat
P+\sum_{k\in\Lambda}\hat O_k\label{64}\en Of course $\hat N$ is
associated to the total number of shares in our closed market and,
therefore, is called the {\em total number operator}. $\hat K$ is
the total amount of money present in the market and is called the
{\em total cash operator}. $\hat \Gamma$ has not a direct
interpretation so far, since is just the sum of the price  and the
{\em total supply} operators, $\hat O=\sum_{k\in\Lambda}\hat O_k$.
It may be worth recalling that the supply operators are only related
to the reservoir $\R$, because of our initial choice. This is the
reason why there is no contribution to the operator $\hat O$ coming
from $\Sc$.

\begin{prop}
The operators $\hat N$, $\hat K$ and $\hat \Gamma$ are constants of
motion.
\end{prop}
The proof of this proposition is a simple exercise based on the
commutation rules above. Indeed, it is not hard to check that $H$
commutes with $\hat N$, $\hat K$ and with $\hat\Gamma$. This proves
that our main motivation for introducing the hamiltonian in
(\ref{61}) is correct: with this choice we are constructing a closed
market in which the total amount of money and the total number of
shares are preserved and in which, if the total supply increases,
then the price of the share must decrease in order for $\hat \Gamma$
to stay constant. Of course, it would be interesting to relate the
changes of $\hat O$ to other (maybe external) conditions, but this
will problem will be considered elsewhere: here we just consider the
simplified point of view for which  $\hat O$ may change in time, but
we don't analyze the reason why this happens.

The next step of our analysis should be to recover the equations of
motion for the portfolio of the trader $\tau$, defined in analogy
with (\ref{511}) as \be \hat \Pi(t)=\hat P(t)\,\hat n(t)+\hat
k(t).\label{65}\en It is not surprising that this cannot be done
exactly so that some perturbative technique is needed. We will
consider in the following sub-sections two {\em orthogonal}
approaches, orthogonal in the sense that they give different
information under different conditions which, together, help in a
better understanding of the model. In particular we will first
consider the so-called {\em stochastic limit} of the system: this
approximation will produce the explicit form of the generator of the
semigroup arising from the hamiltonian (\ref{61}), and this will
give some interesting condition for the {\em stationarity} of the
model, i.e. for $\hat \Pi(t)$ to be  constant in time. We will see
that this is possible under certain conditions on the parameters
defining the model. The second approach will make use of a sort of
FPL approximation which will produce a system of differential
equation for the mean value of $\hat \Pi(t)$ whose solution can be
explicitly found.

\subsection{the stochastic limit of the model}

The stochastic limit of a quantum system is a perturbative strategy
widely discussed in \cite{accbook} and which proved to be quite
useful in the analysis of several quantum mechanical systems, see
\cite{bag2} for a recent review of some  applications of this
procedure to many-body systems.

Here we adopt this procedure {\em pragmatically}, i.e. without
discussing any detail, while, to keep the paper self-contained, we
postpone to Appendix B the list of some basic facts of this
approach.

The first step consists in obtaining the free time evolution of the
interaction hamiltonian which we still call, with a small abuse of
language, $H_I(t)$. Due to the commutation rules (\ref{62}) and
(\ref{63}) we find that \be H_I(t):=e^{iH_0t}H_I e^{-iHt}= z^\dagger
\,Z(f\,e^{it\hat \varepsilon_Z})+z\,Z^\dagger(\overline
f\,e^{-it\hat\varepsilon_Z})+p^\dagger \,o(g\,e^{it\varepsilon_0})+p
\,o^\dagger(\overline g\,e^{-it\varepsilon_0}),
 \label{66}\en
where we have defined \be \hat\varepsilon_Z(k):=\hat
P(\Omega_C(k)-\omega_c)-(\Omega_A(k)-\omega_a),\quad
\varepsilon_O(k):=\omega_p-\Omega_O(k)\label{67}\en and, for
instance,
$Z(f\,e^{it\hat\varepsilon_Z})=\sum_{k\in\Lambda}f(k)\,e^{it\hat\varepsilon_Z(k)}\,Z_k$.

The next step consists in computing first
$\omega\left(H_I\left(\frac{t_1}{\lambda^2}\right)H_I\left(\frac{t_2}{\lambda^2}\right)\right)$,
then
$$I_\lambda(t)=\left(-\frac{i}{\lambda}\right)^2\int_0^t dt_1
\int_0^{t_1}dt_2\,\omega\left(
H_I\left(\frac{t_1}{\lambda^2}\right)H_I\left(\frac{t_2}{\lambda^2}\right)\right),
$$ and finally the limit of $I_\lambda(t)$ for $\lambda\rightarrow 0$. Here $\omega$
is a state of the market, which we take as a product state
$\omega=\omega_{sys}\otimes \omega_{res}$ with $\omega_{sys}$ a
gaussian state, that is it satisfies $\omega_{sys}(a^\sharp)=
\omega_{sys}(c^\sharp)=\omega_{sys}(p^\sharp)=0$ and
$\omega_{sys}(a\,a)=\omega_{sys}(c\,c)=\omega_{sys}(a^\dagger\,a^\dagger)=
\omega_{sys}(p\,p)=\omega_{sys}(p^\dagger\,p^\dagger)=0$. Here
$a^\sharp$ can be $a$ or $a^\dagger$ and the same notation is
adopted for $c^\sharp$ and $p^\sharp$. These conditions are
obviously satisfied if $\omega_{sys}$ is a vector state analogous to
that in (\ref{33}). We don't give here the details of the
computation, which is rather straightforward, but only the final
result which is obtained under the assumptions that the two
functions $\varepsilon_Z(k):=\omega(\hat\varepsilon_Z(k))$ and
$\varepsilon_O(k)$ are not identically zero. Moreover, it is
convenient to assume that \be \varepsilon_Y(k)=\varepsilon_Y(q)
\,\Longleftrightarrow\, k=q, \label{68}\en where $Y=Z,O$. Then, if
we define the following complex constants
\be\left\{\begin{array}{ll}
\Gamma_Z^{(a)}=\sum_{k\in\Lambda}\,|f(k)|^2\,\omega_{res}(Z_k\,Z_k^\dagger)\,\int_{-\infty}^0
\,d\tau\,e^{-i\tau\varepsilon_Z(k)}\\
\Gamma_Z^{(b)}=\sum_{k\in\Lambda}\,|f(k)|^2\,\omega_{res}(Z_k^\dagger\,Z_k)\,\int_{-\infty}^0
\,d\tau\,e^{i\tau\varepsilon_Z(k)}\\
\Gamma_O^{(a)}=\sum_{k\in\Lambda}\,|g(k)|^2\,\omega_{res}(o_k\,o_k^\dagger)\,\int_{-\infty}^0
\,d\tau\,e^{-i\tau\varepsilon_O(k)}\\
\Gamma_O^{(b)}=\sum_{k\in\Lambda}\,|g(k)|^2\,\omega_{res}(o_k^\dagger\,o_k)\,\int_{-\infty}^0
\,d\tau\,e^{i\tau\varepsilon_O(k)}\\
\end{array}\right. \label{69}\en which surely exist if $f(k)$ and $g(k)$ are regular enough,
we get
$$I(t)=-t\left\{\omega_{sys}(z^\dagger\,z)\Gamma_Z^{(a)}+
\omega_{sys}(z\,z^\dagger)\Gamma_Z^{(b)}+\omega_{sys}(p^\dagger\,p)\Gamma_O^{(a)}+
\omega_{sys}(p\,p^\dagger)\Gamma_O^{(b)}\right\}
$$
Next we need to find the expression of a self-adjoint, time
dependent operator $H^{(ls)}(t)$, the so-called {\em stochastic
limit} hamiltonian, which reproduces this result in a sense that we
will specify in a moment.

Let us take \be
H^{(ls)}(t)=z^\dagger\left(Z^{(a)}(t)+{Z^{(b)}}^\dagger(t)\right)+
z\left({Z^{(a)}}^\dagger(t)+{Z^{(b)}}(t)\right)+$$
$$+
p^\dagger\left(o^{(a)}(t)+{o^{(b)}}^\dagger(t)\right)+p\left({o^{(a)}}^\dagger(t)+
{o^{(b)}}(t)\right)\label{610}\en where the new operators introduced
here are assumed to satisfy the following commutation rules: \be
\left[Z^{(a)}(t),{Z^{(a)}}^\dagger(t')\right]=\Gamma_Z^{(a)}\,\delta_+(t-t'),
\quad
\left[Z^{(b)}(t),{Z^{(b)}}^\dagger(t')\right]=\Gamma_Z^{(b)}\,\delta_+(t-t'),
\en and \be
\left[o^{(a)}(t),{o^{(a)}}^\dagger(t')\right]=\Gamma_O^{(a)}\,\delta_+(t-t'),
\quad
\left[o^{(b)}(t),{o^{(b)}}^\dagger(t')\right]=\Gamma_O^{(b)}\,\delta_+(t-t'),
 \en
if $t\geq t'$. The time ordering is crucial here and $\delta_+$ is
essentially the Dirac delta functions but for a normalization which
arises because of the time ordering we consider here,
\cite{accbook}. The only property of $\delta_+$ which we will need
is the following: $\int_0^t\,\delta_+(t-\tau)\,h(\tau)\,d\tau=h(t)$.

Now, let $\Psi_0$ be the vacuum of the operators $Z^{(a)}(t)$,
$Z^{(b)}(t)$, $o^{(a)}(t)$ and $o^{(b)}(t)$. This means that
$Z^{(a)}(t)\Psi_0= Z^{(b)}(t)\Psi_0= o^{(a)}(t)\Psi_0=
o^{(b)}(t)\Psi_0=0$ for all $t\geq 0$. Then, if we consider
$\Omega(.)=\omega_{sys}(.)\otimes <\Psi_0,\,.\,\Psi_0>$ and we
compute
$$
J(t)=(-i)^2\,\int_0^t dt_1 \int_0^{t_1}dt_2\,\Omega\left(
H^{(ls)}(t_1)\,H^{(ls)}(t_2)\right),
$$
we conclude that $J(t)=I(t)$. This means that, at a first order,
$H^{(ls)}(t)$ allows us to get the same wave operator $U_t$ which
describes the time evolution of the systems. We use $H^{(ls)}(t)$ to
construct the wave operator as
$U_t=\1-i\int_0^t\,H^{(ls)}(t')\,U_t'$, and then to deduce the
following commutation rules: \be
\left[Z^{(a)}(t),U_t\right]=-i\Gamma_Z^{(a)}\,z\,U_t, \quad
\left[Z^{(b)}(t),U_t\right]=-i\Gamma_Z^{(b)}\,z^\dagger\,U_t
\label{611}\en and \be
\left[o^{(a)}(t),U_t\right]=-i\Gamma_O^{(a)}\,p\,U_t, \quad
\left[o^{(b)}(t),U_t\right]=-i\Gamma_O^{(b)}\,p^\dagger\,U_t
\label{612}\en by making use of the {\em time consecutive
principle}, \cite{accbook}.

We are now ready to get the expression of the generator. Let $X$ be
a generic {\em observable} of the system, that is, in our present
context, some dynamical variable related to the trader $\tau$. Let
$\1_r$ be the identity operator of the reservoir. Then the time
evolution of $X\otimes\1_r$ in the interaction picture is given by
$j_t(X\otimes\1_r)=U_t^\dagger(X\otimes\1_r)U_t$, so that
$$
\partial_t
j_t(X\otimes\1_r)=iU_t^\dagger[H^{(ls)}(t),X\otimes\1_r]U_t
$$
Using now the commutators in (\ref{611}) and (\ref{612}), and
recalling that $\Psi_0$ is annihilated by all the {\em new}
reservoir operators, we find that $$\Omega\left(\partial_t
j_t(X\otimes\1_r)\right)=\Omega(U_t^\dagger\{\Gamma_Z^{(a)}[z^\dagger,X]\,z-
\overline{\Gamma_Z^{(a)}}z^\dagger\,[z,X]+\Gamma_Z^{(b)}[z,X]\,z^\dagger-
\overline{\Gamma_Z^{(b)}}z\,[z^\dagger,X]+
$$
$$+\Gamma_O^{(a)}[p^\dagger,X]\,p-
\overline{\Gamma_O^{(a)}}p^\dagger\,[p,X]+\Gamma_O^{(b)}[p,X]\,p^\dagger-
\overline{\Gamma_O^{(b)}}p\,[p^\dagger,X]\}U_t)
$$ which, together
with the equality $\Omega\left(\partial_t
j_t(X\otimes\1_r)\right)=\Omega(j_t(L(X\otimes\1_r)))$, gives us the
following expression of the generator: $$
L(X\otimes\1_r)=\Gamma_Z^{(a)}[z^\dagger,X]\,z-
\overline{\Gamma_Z^{(a)}}z^\dagger\,[z,X]+\Gamma_Z^{(b)}[z,X]\,z^\dagger-
\overline{\Gamma_Z^{(b)}}z\,[z^\dagger,X]+$$ \be+
\Gamma_O^{(a)}[p^\dagger,X]\,p-
\overline{\Gamma_O^{(a)}}p^\dagger\,[p,X]+\Gamma_O^{(b)}[p,X]\,p^\dagger-
\overline{\Gamma_O^{(b)}}p\,[p^\dagger,X]\} \label{613}\en Therefore
we find, after few computations, \be L(\hat
n\otimes\1_r)=2\Re\{\Gamma_Z^{(b)}\}z\,z^\dagger-2\Re\{\Gamma_Z^{(a)}\}\,z^\dagger\,z,\label{614}\en
and \be L(\hat k\otimes\1_r)=-2\,\hat
P\,\Re\{\Gamma_Z^{(b)}\}z\,z^\dagger+2\,\hat
P\,\Re\{\Gamma_Z^{(a)}\}\,z^\dagger\,z,\label{615}\en which in
particular shows that $L(\hat k\otimes\1_r)+\hat P\,L(\hat
n\otimes\1_r)=0$. Finally we find, using these results and recalling
that $\hat \Pi(t)=\hat P(t)\,\hat n(t)+\hat k(t)$, \be L(\hat
\Pi\otimes\1_r)=2\,(\Re\{\Gamma_O^{(b)}\}-\Re\{\Gamma_O^{(a)}\})\hat
P\,\hat n+2\,\Re\{\Gamma_O^{(b)}\}\,\hat n.\label{616}\en The first
remark is that, in the stochastic limit, even if the time dependence
of $\hat n$ and $\hat k$ depend on $\Gamma_Z^{(a)}$ and
$\Gamma_Z^{(b)}$, the time evolution of $\hat \Pi$ in a first
approximation do not! In fact,  formula (\ref{616}) shows that it
only depends on $\Gamma_O^{(a)}$ and $\Gamma_O^{(b)}$. However,
since  the time evolution of $\hat n$ depends on $\Gamma_Z^{(a,b)}$
because of (\ref{614}) and (\ref{615}), this dependence will
necessarily play a role also in $\hat\Pi(t)$.

The above equations show that, even after the stochastic limit has
been taken, it is quite difficult to produce a closed set of
differential equations. On the contrary  it is quite easy to deduce
conditions for the stationarity of the market. This is exactly what
we will discuss next.

We begin noticing that, for instance, we have \be
2\Re\{\Gamma_O^{(a)}\}=\sum_{k\in\Lambda}|g(k)|^2
\,\omega_{res}(o_k\,o_k^\dagger)\,\int_{\Bbb{R}}\,d\tau\,e^{-i\tau\,\varepsilon_O(k)}=2\pi
\sum_{k\in\Lambda}|g(k)|^2
\,\omega_{res}(o_k\,o_k^\dagger)\,\delta(\varepsilon_O(k))\label{617}\en
and analogously we find that $ \Re\{\Gamma_O^{(b)}\}=\pi
\sum_{k\in\Lambda}|g(k)|^2
\,\omega_{res}(o_k^\dagger\,o_k)\,\delta(\varepsilon_O(k))$ while
$\Re\{\Gamma_Z^{(a)}\}=\pi \sum_{k\in\Lambda}|f(k)|^2
\,\omega_{res}(Z_k\,Z_k^\dagger)\,\delta(\varepsilon_Z(k))$ and
$\Re\{\Gamma_Z^{(b)}\}=\pi \sum_{k\in\Lambda}|f(k)|^2
\,\omega_{res}(Z_k^\dagger\,Z_k)\,\delta(\varepsilon_Z(k))$.

Therefore, since $[o_k,o_k^\dagger]=\1$ and, as a consequence,
$\omega_{res}(o_k\,o_k^\dagger)-\omega_{res}(o_k^\dagger\,o_k)=\omega_{res}(\1)=1$,
we find that
$\Re\{\Gamma_O^{(b)}\}-\Re\{\Gamma_O^{(a)}\}=-\pi\sum_{k\in\Lambda}|g(k)|^2\delta(\varepsilon_O(k))$.
The conclusion now follows from (\ref{616}): the portfolio of $\tau$
is {\em stationary} (in our approximation) when the function
$\varepsilon_O(k)$ has no zero for $k\in\Lambda$. Indeed, if this is
the case, we deduce that $L(\hat \Pi\otimes\1_r)=0$. Since
$\varepsilon_O(k)=\omega_p-\Omega_O(k)$ this means that if the free
dynamics of the price and the supply are based on substantially
different quantities then the portfolio of $\tau$ keeps its original
value, even if the operators $\hat n(t)$ and $\hat k(t)$ may
separately change with time. This is an interesting result since it
can be summarized just stating that, within the approximation we are
considering here, the fact that $\hat\Pi(t)$ depends or not on time
is only related to a given {\em equilibrium}, if any, between the
free price hamiltonian, $\omega_p\,p^\dagger\,p$, and the free
supply hamiltonian,
$\sum_{k\in\Lambda}\,\Omega_O(k)\,c_k^\dagger\,c_k$: again, the
interplay between these two ingredients of the model play an
interesting role!

\vspace{1mm}

A similar analysis can be carried out also to get  conditions for
the equilibrium of $\hat n(t)$ and $\hat k(t)$. Because of
(\ref{614}) and (\ref{615}), and because of the known time evolution
of $\hat P(t)$, $\hat n(t)$ is constant if and only if $\hat k(t)$
is constant, and for this to be true the function $\varepsilon_Z(k)$
must be different from zero for each $k\in\Lambda$. On the other
hand, if at least one zero of $\varepsilon_Z(k)$ exists in
$\Lambda$, then a non-stationary condition for $\hat n(t)$ and $\hat
k(t)$ is possible.

\subsection{a different approximation}

The approach we have discussed so far produced some interesting
information about the stationarity of the portfolio of $\tau$ but no
concrete insight about its time evolution. In other words, if we try
to deduce the time behavior of $\hat\Pi(t)$ we get no significant
simplification if we adopt the form of the generator in (\ref{613})
or if we look directly to the Heisemberg expression for
$\hat\Pi(t)$, $\hat\Pi(t)=e^{iHt}\,\hat\Pi(0)\,e^{-iHt}$. However,
also this last attempt does not produce directly a closed system of
differential equations: some different approximation must be
assumed. This different approximation will be discussed in this
subsection.

We first remind that given a generic operator $X$ its time
evolution, in the Heisemberg representation, is (formally) given by
$X(t)=e^{iHt}\,X\,e^{-iHt}$, and it satisfies the following
Heisemberg equation of motion: $\dot X(t)=ie^{iHt}[H,X]e^{-iHt}=
i[H,X(t)]$. In the attempt of deducing the analytic expression for
$\hat\Pi(t)$, the following differential equations can be deduced:
\be\left\{\begin{array}{lll}\frac{d\hat
n(t)}{dt}=i\lambda\left(-z^\dagger(t)\,Z(f,t)+z(t)\,Z^\dagger(\overline
f,t)\right),\\
\frac{d\hat k(t)}{dt}=i\lambda\,\hat
P(t)\,\left(z^\dagger(t)\,Z(f,t)-z(t)\,Z^\dagger(\overline
f,t)\right),\\
\frac{d\hat P(t)}{dt}=i\lambda\left(p(t)\,o^\dagger(\overline
g,t)-p^\dagger(t)\,o(g,t)\right),\\
\frac{dz(t)}{dt}=i\left(\hat P(t)\omega_c-\omega_a\right)\,z(t)+i\lambda[z^\dagger(t),z(t)]\,Z(f,t),\\
\frac{dZ(f,t)}{dt}=i\,Z\left((\hat
P(t)\Omega_C-\Omega_A)f,t\right)+i\lambda\,z(t)\,[Z^\dagger(\overline
f,t),
Z(f,t)].\\
\end{array}\right.\label{618}\en
where we have defined $Z(f,t):=e^{iHt}Z(f)e^{-iHt}$, $Z\left((\hat
P(t)\Omega_C-\Omega_A)f,t\right)=\sum_{k\in\Lambda}(\hat
P(t)\Omega_C(k)-\Omega_A(k))\,f(k)\,Z_k(t)$,
$o(g,t)=e^{iHt}\,o(g)\,e^{-iHt}$, and so on.

It is clear that the system (\ref{618}) is not closed, since for
instance the differential equation for $\hat P(t)$ involves $p(t)$,
$o(g,t)$ and their adjoint. This is not a major problem since, as in
Sections II and III and in \cite{bag1}, it is quite easy to deduce
the time evolution of the price operator $\hat P$ with no
approximation at all. This is because $p(t)$ (and $o(g,t)$) can be
found explicitly. Even if these operators can be found under more
general conditions, we will now restrict the model requiring that
the coefficients in $H$ satisfy some extra requirement, which are
only useful to simplify the computations. For instance, we will
assume that $\Omega_O(k)$ is constant in $k$, $\Omega_O=\Omega_O(k)$
for all $k\in\Lambda$, and that
$\omega_p=\sum_{k\in\Lambda}|g(k)|^2=\lambda=\Omega_O$. Then we get
$p(t)=\frac{1}{2}\left(p(e^{-2i\lambda t}+1)+o(g)(e^{-2i\lambda
t}-1)\right)$ and $\hat P(t)=p^\dagger(t)\,p(t)$. Since $\hat P(t)$
depends only on the operators $p$ and $o$, and not on $a, c,$ and so
on, and since we  are interested to the mean value of the operators
in (\ref{618}) in a vector state $\omega$ generalizing (\ref{33}),
we replace this system with its {\em semiclassical} approximation
\be\left\{\begin{array}{lll}\frac{d\hat
n(t)}{dt}=i\lambda\left(-z^\dagger(t)\,Z(f,t)+z(t)\,Z^\dagger(\overline
f,t)\right),\\
\frac{d\hat k(t)}{dt}=i\lambda\,
P_c(t)\,\left(z^\dagger(t)\,Z(f,t)-z(t)\,Z^\dagger(\overline
f,t)\right),\\
\frac{dz(t)}{dt}=i\left( P_c(t)\omega_c-\omega_a\right)\,z(t)+i\lambda[z^\dagger(t),z(t)]\,Z(f,t),\\
\frac{dZ(f,t)}{dt}=i\,Z\left((
P_c(t)\Omega_C-\Omega_A)f,t\right)+i\lambda\,z(t)\,[Z^\dagger(\overline
f,t),
Z(f,t)],\\
\end{array}\right.\label{619}\en
where \be P_c(t)=\omega(\hat
P(t))=\frac{1}{2}\left[(M+O)+(M-O)\cos(2\lambda t)\right]
\label{620}\en  We refer to \cite{bag1} for a more complete
discussion of the two-fold role of the state $\omega$. Here we just
want to remark that a given vector state allows us to pass from the
{\em quantum} dynamics of the model to its {\em classical}
counterpart, since we use $\omega$ to replace the time dependent
operators with their mean values, which are functions of time. At
the same time, moreover, a vector state is used to fix the initial
conditions of the differential equations, that is the initial number
of shares, the initial cash and so on.

In order to simplify further the analysis of this system, it is also
convenient to assume that both $\Omega_C(k)$ and $\Omega_A(k)$ are
constant for $k\in\Lambda$. Indeed, under this assumption, the last
two equation in (\ref{619}) forms by themselves a closed system of
differential equations in the non abelian variables $z(t)$ and
$Z(f,t)$: \be\left\{\begin{array}{lll}
\frac{dz(t)}{dt}=i\left( P_c(t)\omega_c-\omega_a\right)\,z(t)+i\lambda\,Z(f,t)\,[z^\dagger(t),z(t)],\\
\frac{dZ(f,t)}{dt}=i\,(P_c(t)\Omega_C-\Omega_A)\,Z(f,t)+i\lambda\,z(t)\,[Z^\dagger(\overline
f,t),Z(f,t)].\\
\end{array}\right.\label{621}\en

Getting the exact solution of the system (\ref{619}), with
(\ref{621}) as the two last equations, is an hard job. However, this
is a good starting point for finding an approximated solution of the
dynamical problem. Indeed, a natural approach consists in taking the
first non trivial contribution of the system, as usually done in
perturbation theory. This means that, in system (\ref{621}), the
contributions containing the commutators must be neglected since
they are proportional to $\lambda$ while $i\left(
P_c(t)\omega_c-\omega_a\right)\,z(t)$ and
$i\,(P_c(t)\Omega_C-\Omega_A)\,Z(f,t)$ which, on the other way do
not depend on $\lambda$, give a relevant contribution. On the other
way, in order not to trivialize the system, we have to keep the
first two equations in (\ref{619}) as they are: if we simply put
$\lambda=0$ here, in fact, we would trivialize the time evolution of
both $\hat n(t)$ and $\hat k(t)$. With this choice we get
\be\left\{\begin{array}{lll}\frac{d\hat
n(t)}{dt}=i\lambda\left(-z^\dagger(t)\,Z(f,t)+z(t)\,Z^\dagger(\overline
f,t)\right),\\
\frac{d\hat k(t)}{dt}=i\lambda\,
P_c(t)\,\left(z^\dagger(t)\,Z(f,t)-z(t)\,Z^\dagger(\overline
f,t)\right),\\
\frac{dz(t)}{dt}=i\left( P_c(t)\omega_c-\omega_a\right)\,z(t),\\
\frac{dZ(f,t)}{dt}=i\,\left(
P_c(t)\Omega_C-\Omega_A\right)\,Z(f,t).\\
\end{array}\right.\label{622}\en
However, we will now show that this approximation is too rude,
meaning with this that, even if the operators $\hat n(t)$ and $\hat
k(t)$ have a non trivial dynamics, at the classical level we deduce
that both $n(t)=\omega(\hat n(t))$ and $k(t)=\omega(\hat k(t))$ are
constant in time, so that the time behavior of the portfolio
$\Pi(t)=P_c(t)\,n(t)+k(t)=P_c(t)\,n+k$ is uniquely given by
$P_c(t)$.

We first observe that $z(t)$ and $Z(f,t)$ in (\ref{622}) are \be
z(t)=z\,e^{i\chi(t)},\qquad
Z(f,t)=Z(f)\,e^{i\tilde\chi(t)},\label{623}\en where \be
\chi(t)=\alpha t+\beta\sin(2\lambda t),\qquad
\tilde\chi(t)=\tilde\alpha t+\tilde\beta\sin(2\lambda t)
\label{624}\en with \be\left\{\begin{array}{lll}
\alpha=\frac{1}{2}((M+O)\omega_c-2\omega_a),\qquad
\beta=\frac{\omega_c}{4\lambda}(M-O)\\
\tilde\alpha=\frac{1}{2}((M+O)\Omega_C-2\Omega_A),\qquad
\tilde\beta=\frac{\Omega_C}{4\lambda}(M-O)
\label{625}\end{array}\right.\en Our claim is now an immediate
consequence of (\ref{623}) above. Indeed, from the first equation in
(\ref{622}),  taking its mean value on the number vector state
$\omega$ we find $$\dot n(t)=\frac{d}{dt}\,\omega(\hat
n(t))=\omega\left(\frac{d}{dt}\hat
n(t)\right)=i\lambda\left\{-\omega\left(z^\dagger(t)Z(f,t)\right)+
\omega\left(z(t)Z^\dagger(\overline{f},t)\right)\right\}=0$$since,
for instance,
$\omega\left(z^\dagger(t)Z(f,t)\right)=e^{-i(\chi(t)-\tilde\chi(t))}\omega(z^\dagger
Z(f))=0$. Analogously we find that $\dot
k(t)=\frac{d}{dt}\,\omega(\hat k(t))=0$. Therefore we have $n(t)=n$
and $k(t)=k$, as claimed above.

\vspace{2mm}

{\bf Remark:} in a certain sense this result relates the two
approximations considered so far. Indeed, replacing (\ref{619}) with
(\ref{622}) we obtain a stationary behavior for $n(t)$ and $k(t)$.
An analogous behavior was deduced, in the previous subsection, if
$\varepsilon_Z(k)$ has no zero. However, these two different
approximations cannot be directly compared. The reason is the
following: in the stochastic limit approach we need to require that
$\varepsilon_Z(k)$ and $\varepsilon_O(k)$ are not identically zero.
This is crucial to ensure the existence of
$\lim_{\lambda,0}\,I_\lambda(t)$. In the present approximation we
are  requiring that both $\Omega_C(k)$ and $\Omega_A(k)$ are
constant in $k$ so that, see (\ref{67}), we would get
$\varepsilon_Z(k)=P_c(t)(\Omega_C-\omega_c)-(\Omega_A-\omega_a)$,
which may have some zero in $k$ only if it is identically zero in
$k$. In other words, in the conditions in which we are working here
the stochastic limit approach does not work. Viceversa, if we are in
the assumptions of the previous subsection, then system (\ref{619})
cannot be easily solved! Hence the two approximations cover
different situations.

\vspace{2mm}

A better approximation can be constructed. Again the starting point
is the system (\ref{621}), for which we now construct iteratively a
solution, stopping at the first relevant order. In other words, we
take $z_0(t)$ and $Z_0(f,t)$ as in (\ref{623}),
$z_0(t)=z\,e^{i\chi(t)}$ and $Z_0(f,t)=Z(f)\,e^{i\tilde\chi(t)}$,
and then we look for the next approximation by considering the
following system:
$$\left\{\begin{array}{lll}
\frac{dz_1(t)}{dt}=i\left( P_c(t)\omega_c-\omega_a\right)\,z_0(t)+i\lambda\,Z_0(f,t)\,[z_0^\dagger(t),z_0(t)],\\
\frac{dZ_1(f,t)}{dt}=i\,(P_c(t)\Omega_C-\Omega_A)\,Z_0(f)+i\lambda\,z_0(t)\,[Z_0^\dagger(\overline
f,t),Z_0(f,t)].\\
\end{array}\right.$$
which can be still written as \be\left\{\begin{array}{lll}
\frac{dz_1(t)}{dt}=
i\left( P_c(t)\omega_c-\omega_a\right)\,z_0(t)+i\lambda\,Z_0(f,t)\,[z_0^\dagger,z_0],\\
\frac{dZ_1(f,t)}{dt}=i\,(P_c(t)\Omega_C-\Omega_A)\,Z_0(f)+i\lambda\,z_0(t)\,[Z_0^\dagger(\overline
f),Z_0(f)].\\
\end{array}\right.\label{626}\en
These equations can be solved and the solution can be written as \be
z_1(t)=z\,\eta_1(t)+Z(f)\,[z^\dagger,z]\,\eta_2(t),\quad
Z_1(f,t)=Z(f)\,\tilde\eta_1(t)+z\,[Z(\overline
f)^\dagger,Z(f)]\,\tilde\eta_2(t), \label{627}\en where we have
introduced the following functions \be\left\{\begin{array}{ll}
\eta_1(t)=1+i\int_0^t(
P_c(t')\omega_c-\omega_a)\,e^{i\chi(t')}\,dt',\quad
\eta_2(t)=i\lambda\int_0^te^{i\tilde\chi(t')}\,dt'\\
\tilde\eta_1(t)=1+i\int_0^t(
P_c(t')\Omega_C-\Omega_A)\,e^{i\tilde\chi(t')}\,dt',\quad
\tilde\eta_2(t)=i\lambda\int_0^te^{i\chi(t')}\,dt'
\end{array}\right.\label{627bis}
\en It is not a big surprise that  this approximated solution does
not share with $z(t)$ and $Z(f,t)$ all their properties. In
particular, while for instance $[z(t),Z(f,t)]=0$ for all $t$,
$z_1(t)$ and $Z_1(f,t)$ do not commute. For this reason we consider
the equations for $\hat n(t)$ and $\hat k(t)$ as in (\ref{622}) as
far as possible, replacing $z(t)$ and $Z(f,t)$ with $z_1(t)$ and
$Z_1(f,t)$ only at the last step.

It is easy to find that the mean values of the first two equations
in (\ref{622}) can be written as \be\left\{\begin{array}{lll}\dot
n(t)=\frac{d
n(t)}{dt}=-2\lambda\Im\left\{\omega\left(z(t)\,Z^\dagger(\overline
f,t)\right)\right\},\\
\dot k(t)=\frac{d
k(t)}{dt}=2\lambda\,P_c(t)\Im\left\{\omega\left(z(t)\,Z^\dagger(\overline
f,t)\right)\right\},\\
\end{array}\right.\label{628}\en
which in particular implies a well known identity: $P_c(t)\dot
n(t)+\dot k(t)=0$ for all $t$, which in turns implies that
$\dot\Pi(t)=\dot P_c(t)\,n(t)$. It should be remarked that, because
of this relation, since $M=O$ implies $P_c(t)=P_c(0)=M$, then when
$M=O$ the dynamics of the portfolio of $\tau$ is trivial,
$\Pi(t)=\Pi(0)$, even if both $n(t)$ and $k(t)$ may change in time.

It is now at this stage that we insert $z_1(t)$ and $Z_1(f,t)$ in
the differential equations. If $\omega$ is the usual number state,
and if we call for simplicity
\be\left\{\begin{array}{lll}\omega(1):=\omega\left(zz^\dagger\,[Z^\dagger(\overline
f),Z(f)]\right\},\\
\omega(2):=\omega\left(Z(f)Z^\dagger(\overline
f)\,[z^\dagger,z]\right\},\\
r(t)=\omega(1)\,\eta_1(t)\,\overline{\tilde\eta_2(t)}+\omega(2)\,\eta_2(t)\,\overline{\tilde\eta_1(t)}\\
\end{array}\right.\label{629}\en
then we get \be\left\{\begin{array}{lll} n(t)=n-2\,\lambda\Im\left\{\int_0^t r(t')\,dt'\right\},\\
k(t)=k+2\,\lambda\Im\left\{\int_0^t P_c(t')\,r(t')\,dt'\right\}.\\
\end{array}\right.\label{630}\en

The time dependence of the portfolio can now be written as \be
\Pi(t)=\Pi(0)+\delta\Pi(t),\label{631}\en with $$
\delta\Pi(t)=n(O-M)\sin^2(\lambda t)+$$
\be+\left(-2\lambda\,\Im\left\{\int_0^t
r(t')\,dt'\right\}\left(M+(O-M)\sin^2(\lambda
t)\right)+2\lambda\,\Im\left\{\int_0^t
P_c(t')\,r(t')\,dt'\right\}\right), \label{632}\en which gives the
variation of the portfolio of $\tau$ in time. We observe that, as it
is expected, $\delta\Pi(t)=0$ if $\lambda=0$.  \vspace{2mm}

In the last part of this subsection we look for particular solutions
of this system under special conditions. A more detailed analysis of
these results will be discussed in another paper, which is now in
preparation and where a more general settings will be considered. A
first remark concerning (\ref{632}) is the following: if $O>M$, it
is more likely for $\tau$ to have a positive $\delta\Pi(t)$ if the
number of the shares $n$ in his portfolio at time $t=0$ is large: if
at $t=0$ the supply of the market is larger than the price of the
share then {\em for a trader with many shares it is easier to become
even richer}! If, on the contrary, $O<M$, having a large number of
shares does not automatically produce an increment of the portfolio.

Coefficients $\omega(1)$ and $\omega(2)$ can be found explicitly and
depend on the initial conditions of the market. If, for simplicity's
sake, we consider $\Lambda=\{k_o\}$, that is if  the reservoir
consists of just  another trader interacting with $\tau$, then we
get
$$\omega(1)=|f(k_o)|^2\,(1+n)\,k^{\{-M\}}\left(n'\,k_o^{\{+M\}}-(1+n')\,k_o^{\{-M\}}\right)$$
and
$$\omega(2)=|f(k_o)|^2\,(1+n')\,k_o^{\{-M\}}\left(n\,k^{\{+M\}}-(1+n)\,k^{\{-M\}}\right).$$
It is clear that these coefficients coincide if $k=k_o$ and $n=n'$.

Let us first fix $M=1$, $O=2$, $\lambda=1$, $\omega_a=\omega_c=1$,
$\Omega_A=\Omega_C=2$. Then the plots of $\delta\Pi(t)$ below, in
which $n$ is fixed to be 10, are related to the following different
values of $\omega(1)$ and $\omega(2)$: $(\omega(1),\omega(2))=(1,1),
(1,10), (10,1)$.

\begin{center}
\mbox{\includegraphics[height=3.2cm, width=4.5cm]
{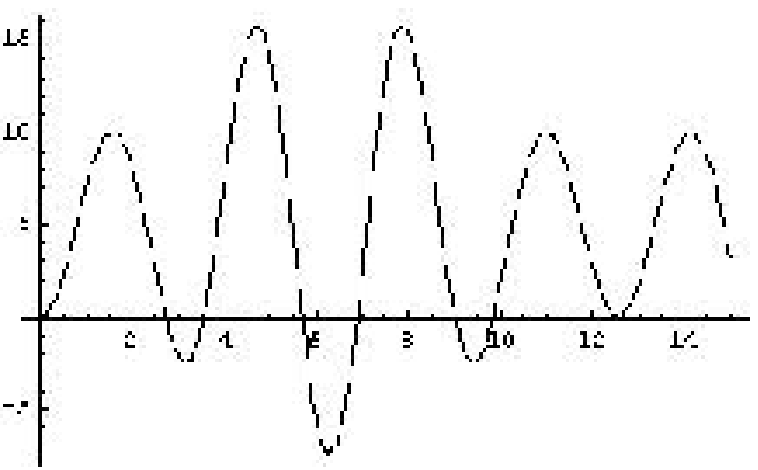}}\hspace{6mm} \mbox{\includegraphics[height=3.2cm,
width=4.5cm] {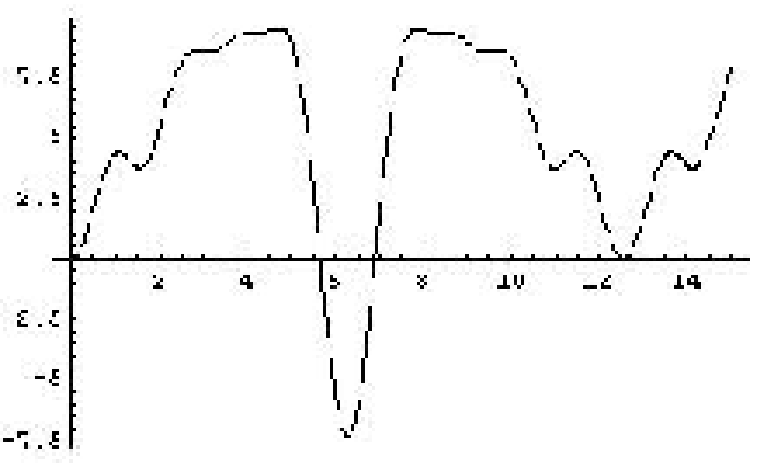}} \hspace{6mm}
\mbox{\includegraphics[height=3.2cm, width=4.5cm] {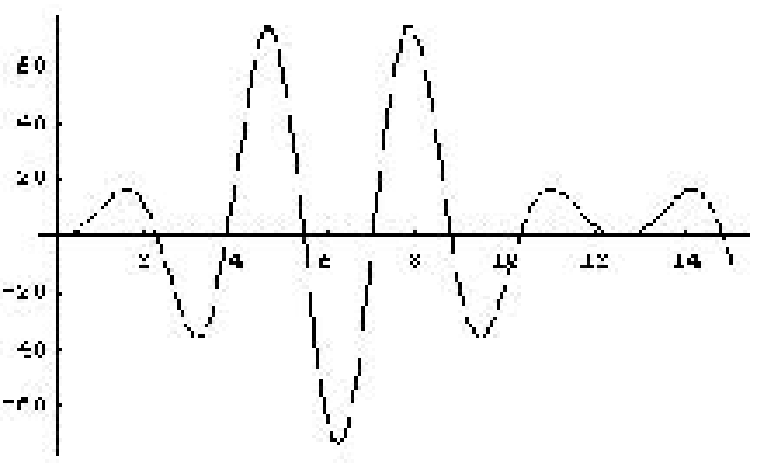}}\hfill\\
\begin{figure}[h]
\caption{\label{fig1}$\delta\Pi(t)$  for $n=10$ and
$(\omega(1),\omega(2))=(1,1)$ (left), $(\omega(1),\omega(2))=(1,10)$
(middle), $(\omega(1),\omega(2))=(10,1)$ (right) }
\end{figure}
\end{center}

The plots do not change much if we fix $n=5$ and, surprisingly
enough, also the ranges of variations of $\delta\Pi(t)$ essentially
coincide with those above: $n$ seems to play no crucial role here!
In Figure 2 we plot $\delta\Pi(t)$ in the same conditions as before,
but for $n=5$.

\begin{center}
\mbox{\includegraphics[height=3.2cm, width=4.5cm]
{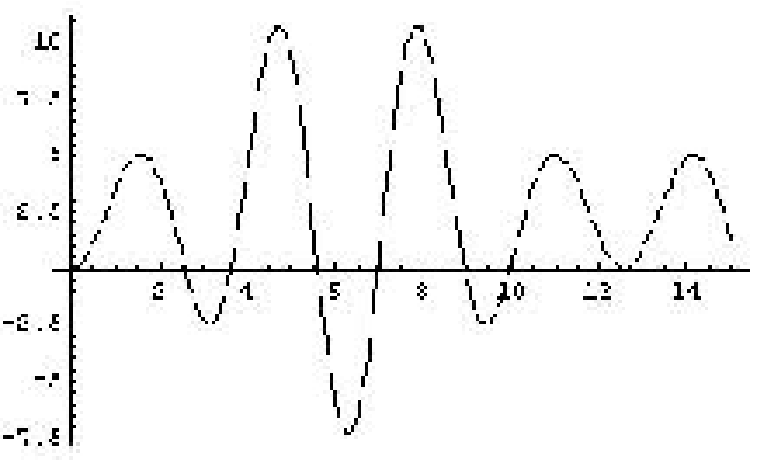}}\hspace{6mm} \mbox{\includegraphics[height=3.2cm,
width=4.5cm] {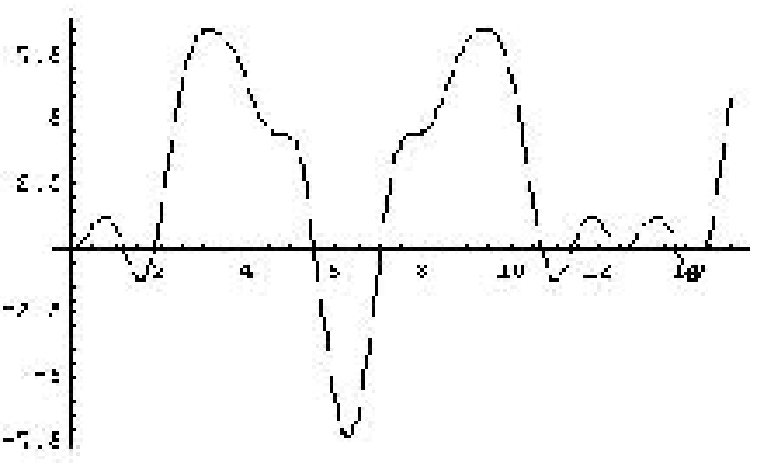}} \hspace{6mm}
\mbox{\includegraphics[height=3.2cm, width=4.5cm] {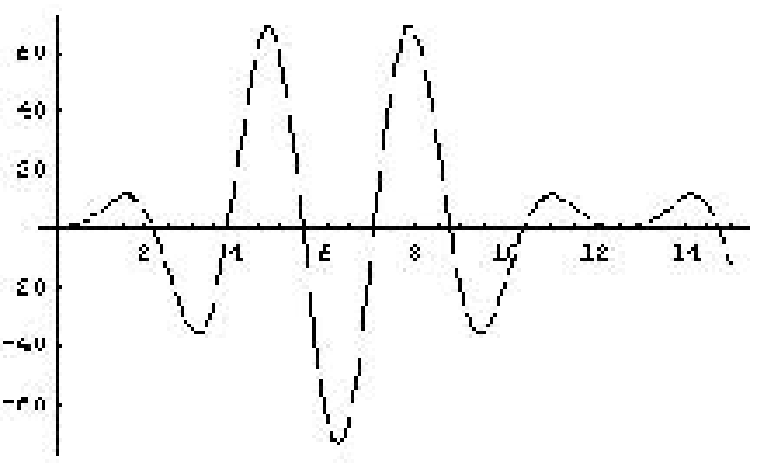}}\hfill\\
\begin{figure}[h]
\caption{\label{fig2}$\delta\Pi(t)$  for $n=5$ and
$(\omega(1),\omega(2))=(1,1)$ (left), $(\omega(1),\omega(2))=(1,10)$
(middle), $(\omega(1),\omega(2))=(10,1)$ (right) }
\end{figure}
\end{center}
From both these figures we see that, for trader $\tau$, the most
convenient situation is $(\omega(1),\omega(2))=(1,10)$: in this case
there is only a small range of time in which $\delta\Pi(t)$ is
negative. For all other times $\delta\Pi(t)$ is positive and
$\Pi(t)$ increases its original value. The situation is a bit less
favorable for other choices of $(\omega(1),\omega(2))$. This is not
surprising since $\omega(1)$ and $\omega(2)$ are related to the
initial values of the stock market we are considering and, how it is
well known, different initial conditions may correspond to quite
different dynamical behaviors!

\vspace{2mm}

Now we change the relation between $M$ and $O$. Therefore we fix
$M=2$, $O=1$, $\lambda=1$, $\omega_a=\omega_c=1$,
$\Omega_A=\Omega_C=2$. Again the plots of $\delta\Pi(t)$ below are
related to the following values of: $(\omega(1),\omega(2))=(1,1),
(1,10), (10,1)$, and we fix $n=10$.

\begin{center}
\mbox{\includegraphics[height=3.2cm, width=4.5cm]
{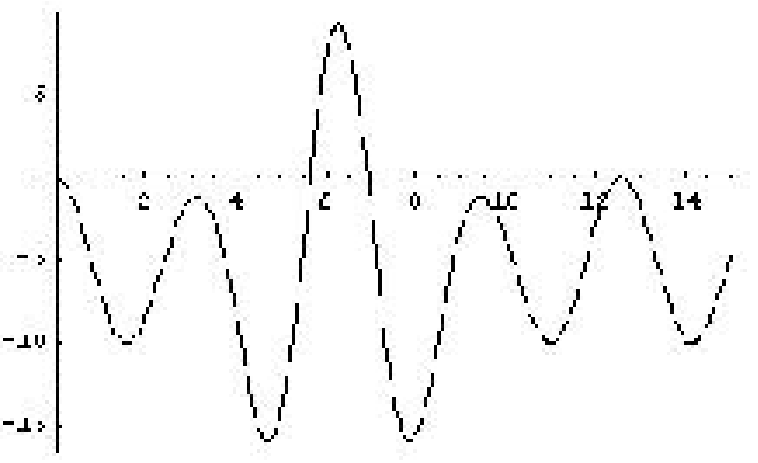}}\hspace{6mm} \mbox{\includegraphics[height=3.2cm,
width=4.5cm] {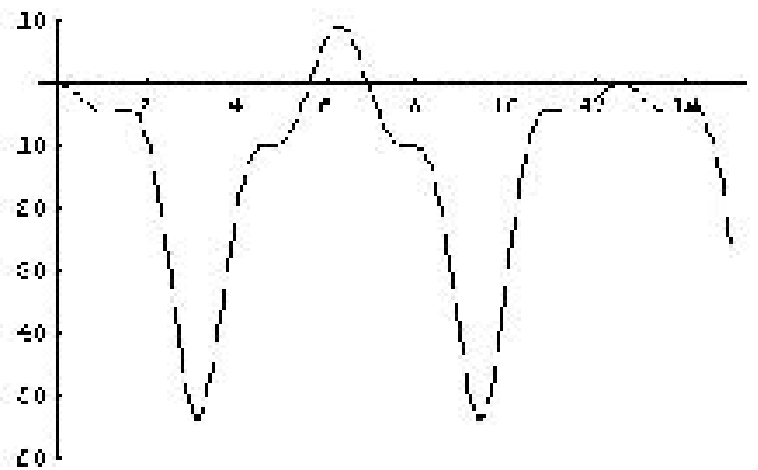}} \hspace{6mm}
\mbox{\includegraphics[height=3.2cm, width=4.5cm] {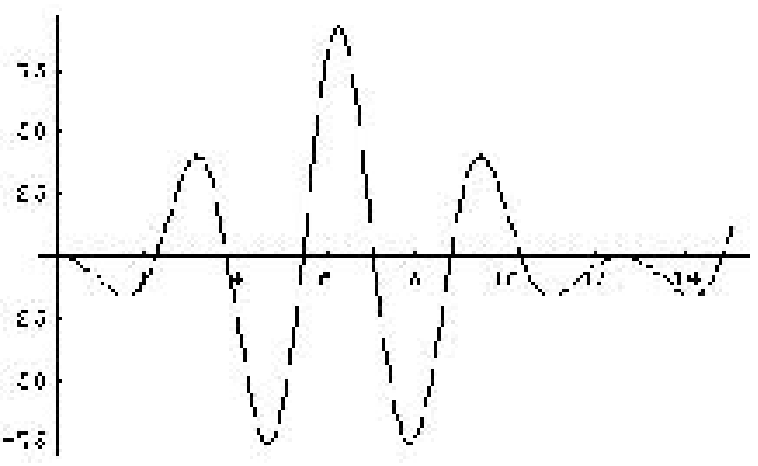}}\hfill\\
\begin{figure}[h]
\caption{\label{fig3}$\delta\Pi(t)$  for $n=10$
$(\omega(1),\omega(2))=(1,1)$ (left), $(\omega(1),\omega(2))=(1,10)$
(middle), $(\omega(1),\omega(2))=(10,1)$ (right) }
\end{figure}
\end{center}
We see that these plots look very much as those in Figure 1 reflexed
with respect to the horizontal axis. This means that, for $n=10$,
the main contribution in (\ref{632}) is the term
$n(O-M)\sin^2(\lambda t)$. Of course this is even more evident if
$n$ is larger than 10, while for small values of $n$ the role of the
other contributions in (\ref{632}) is in general more relevant.

We have already stressed that, if $M=0$, then $\delta\Pi(t)=0$ for
all $t\geq0$. Therefore we don't plot $\delta\Pi(t)$ in this
condition. Instead of this, we finish considering what happens if we
change the values of $\omega_a$ and $\omega_c$ with $\Omega_A$ and
$\Omega_C$. For that we fix, as in the first case, $M=1$, $O=2$,
$\lambda=1$, while we take $\omega_a=\omega_c=2$ and
$\Omega_A=\Omega_C=1$. The related plots are

\begin{center}
\mbox{\includegraphics[height=3.2cm, width=4.5cm]
{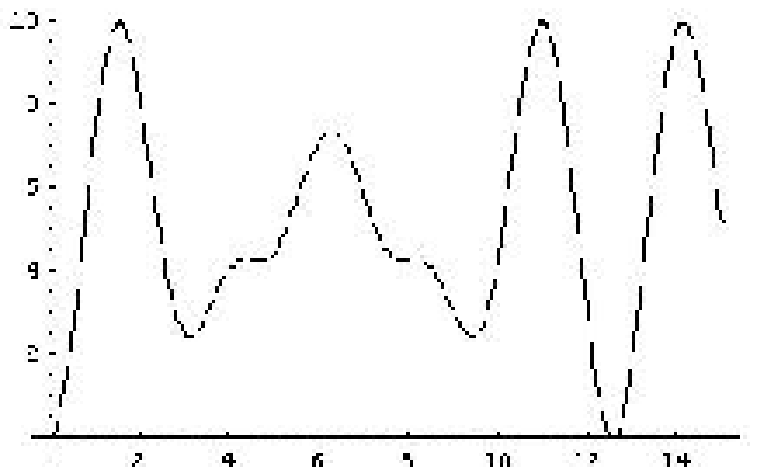}}\hspace{6mm} \mbox{\includegraphics[height=3.2cm,
width=4.5cm] {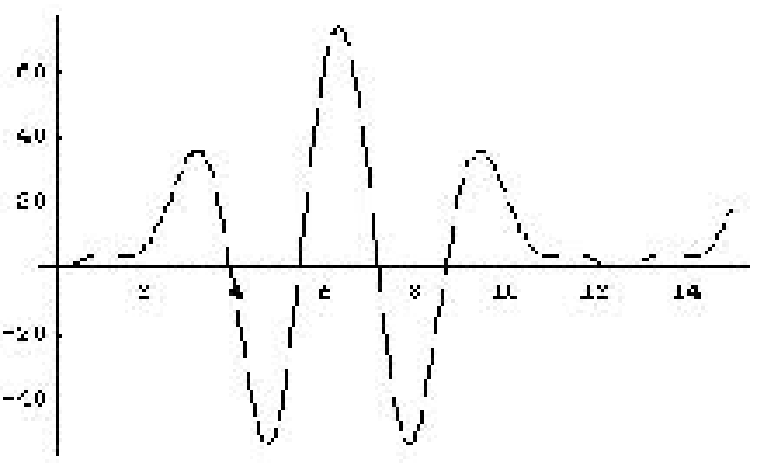}} \hspace{6mm}
\mbox{\includegraphics[height=3.2cm, width=4.5cm] {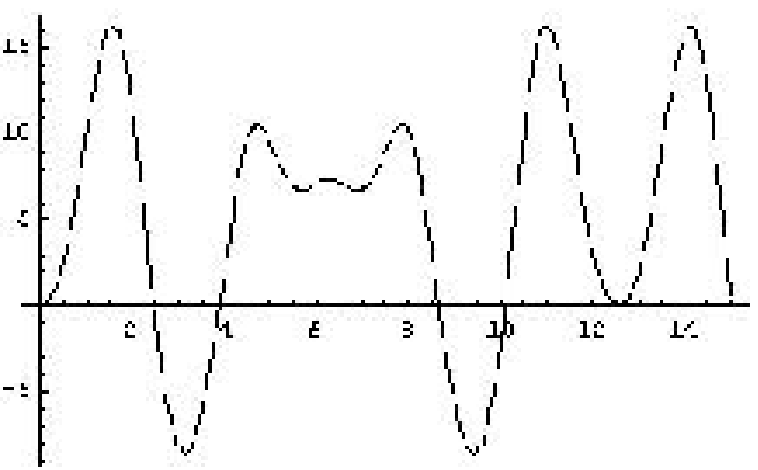}}\hfill\\
\begin{figure}[h]
\caption{\label{fig4}$\delta\Pi(t)$  for $n=10$
$(\omega(1),\omega(2))=(1,1)$ (left), $(\omega(1),\omega(2))=(1,10)$
(middle), $(\omega(1),\omega(2))=(10,1)$ (right) }
\end{figure}
\end{center}

This result is particularly interesting since it shows that, if
$\omega(1)=\omega(2)=1$ and also for $n$ small enough, trader $\tau$
can only improve the value of his portfolio, no matter the value of
$t$. Remember now that $\omega(1)=\omega(2)$ is true if the initial
conditions for the trader $\tau$ and for the trader of the
reservoir, $\sigma$, coincide. Therefore, this suggests that the
relation between the parameters $\omega_a, \omega_c$ and $\Omega_A$,
$\Omega_C$ is crucial to determine $\Pi(t)$ and in particular that
if we take $\Omega_A=\Omega_C>\omega_a=\omega_c$, $\tau$ is, in a
certain sense, in a better condition with respect to trader
$\sigma$. It is therefore natural to associate these parameters, for
instance, to a sort of {\em information} reaching the traders, in
analogy with the interpretation discussed in  \cite{bag1}. We avoid
details here since a deeper analysis is needed for a better
understanding of the role of all the parameters in $H$.  In a
forthcoming paper we will focus our interest exactly on this point:
we will discuss the solution of system (\ref{622}) under several
different conditions, and we will also consider different
expressions of $P_c(t)$, arising from some different {\em
economically reasonable} hamiltonian or by  experimental data.

\section{Conclusions and outcome}

In this paper we have carried on the analysis of a stock market in
terms of Heisemberg dynamics which we  began in \cite{bag1}. In
particular, we have generalized the model introduced in \cite{bag1}
by  introducing a {\em real dynamical behavior} for the price of the
shares. This is, in our opinion, a big achievement with respect to
our previous results.

Section III is just a pedagogical two-traders model which is useful
to fix some general ideas and giving some definitions. The same
model is further generalized in Section IV where a non trivial
market has been introduced. We have considered two different
approximations of this model. The first approximation, the so-called
stochastic limit approach, is useful to get conditions for the
staticity of the portfolio of a given trader. The second
approximation, more useful for the analysis of the general time
evolution of the portfolio, produces many results and is quite
interesting in view of future applications.

In particular, in a close future we plan to add more kind of shares
within the model, and to use system (\ref{619}) with different
functions $P_c(t)$, deduced from other hamiltonian models or by
experimental data. A more long-distance program also includes, for a
market with more shares, an analysis of the role of the Heisenberg
dynamics in the analysis of stock markets. This will be undertaken
clearly  via a comparison between our results and the experimental
data.

We would also like to comment that, as already briefly discussed in
\cite{bagbolo}, the same general strategy seems of some utilities in
contexts which are apparently very far from stock markets and
particle physics. Indeed, the mechanism analyzed here is natural
whenever we are interested in describing exchanges between different
{\em active components} of our (physical, biological,
economical,...) system. Indeed, this is exactly the original remark
which produced second quantization in elementary particle physics,
\cite{mer}. Just as a different example, we may also use the
hamiltonian in (\ref{61}), or some modification of this,  for a
predator-prey system. In this case $\hat n$ represents the {\em
number of predator} operator while $\hat k$ is the {\em number of
prey} operator. The mechanism in $H_I$ implies that when the number
of predator increases of one unit the number of preys decreases of
$<\hat P>$ units, and $\hat P$ can now be interpreted as a sort of
{\em ability} of the predator to catch its victims. We refer to
\cite{bagbolo}, and to a paper in preparation \cite{bagsoc}, for
more applications to sociological contexts.

%\newpage
\section*{Acknowledgements}

This work has been financially supported in part by M.U.R.S.T.,
within the  project {\em Problemi Matematici Non Lineari di
Propagazione e Stabilit\`a nei Modelli del Continuo}, coordinated by
Prof. T. Ruggeri.

It is also my pleasure to thank Dr. Vincenzo Sciacca for his hints
for the numerical results of Section IV.

 \appendix

\renewcommand{\theequation}{A.\arabic{equation}}

 \section{\hspace{-.7cm}ppendix A:  the definition of $c^{\,\hat P}$}

In this Appendix we will discuss in detail how to define the
operators $c^{\,\hat P}$ and ${c^\dagger}^{\,\hat P}$, and some
useful formulas related to them.

To make the situation simpler, we just neglect  here the role of the
other operators appearing in Section III, i.e. the number of share
 and the supply operators, since they play no role in the definition of, say,  $c^{\,\hat P}$.
 We will just consider two sets of bosonic operators $c$ and $p$,
 with $[c,c^\dagger]=[p,p^\dagger]=\1$, and the common vacuum vector
 $\varphi_0$: $c\varphi_0=p\varphi_0=0$. In a standard fashion we call
\be
\varphi_{k,m}=\frac{1}{\sqrt{k!\,m!}}\,(c^\dagger)^k\,(p^\dagger)^m\,\varphi_0,
\label{app1}\en where $k,m\geq 0$. It is well known, \cite{mer},
that $\varphi_{k,m}$ is an eigenstate of $\hat k=c^\dagger c$ and
$\hat P=p^\dagger p$: $\hat k\,\varphi_{k,m}=k\varphi_{k,m}$ and
$\hat P\,\varphi_{k,m}=m\varphi_{k,m}$. Since we have, if $k$ is
large enough, $c\varphi_{k,m}=\sqrt{k}\varphi_{k-1,m}$,
$c^2\varphi_{k,m}=\sqrt{k(k-1)}\varphi_{k-2,m}$, and so on, it is
natural to define \bea c^{\,\hat P}\,\varphi_{k,m}:= \left\{
\begin{array}{ll}
\varphi_{k,m},  &\mbox{ if } m=0,\,\forall k\geq 0;  \\
0,  &\mbox{ if } m>k,\,\forall k\geq 0;\\
\sqrt{k(k-1)\cdots(k-m+1)}\,\varphi_{k-m,m},  & \mbox{ if } k\geq m
> 0
\end{array}
\right. \label{app2}\ena

Analogously, since
$c^\dagger\varphi_{k,m}=\sqrt{k+1}\varphi_{k+1,m}$,
$(c^\dagger)^2\varphi_{k,m}=\sqrt{(k+1)(k+2)}\varphi_{k+2,m}$, and
so on, for all $k$ and $m\geq 0$, we put \bea {c^\dagger}^{\,\hat
P}\,\varphi_{k,m}:= \left\{
\begin{array}{ll}
\varphi_{k,m},  &\mbox{ if } m=0,\,\forall k\geq 0;  \\
\sqrt{(k+1)(k+2)\cdots(k+m)}\,\varphi_{k+m,m},  & \mbox{ if } m>0,
\
\end{array}
\right. \label{app3}\ena

\vspace{2mm}

\noindent{\bf Remark:} We could use a different name for the
operators $c^{\,\hat P}$ and  ${c^\dagger}^{\,\hat P}$. For instance
we could call $\hat Y$ and $\hat W$ the operators defined as in
(\ref{app2}) and  (\ref{app3}): however we have decided to keep this
notation to stress the role of {\bf both} $c$ {\bf and} $\hat P$ in
the definition of these ladder operators.

\vspace{2mm}

These definitions, other than natural, have two nice consequences:
(i) they really define the operators ${c}^{\,\hat P}$ and
${c^\dagger}^{\,\hat P}$ since they are now defined on the vectors
of an orthonormal basis in the Hilbert-Fock space $\Hil$ of the
system, which is the closure of the linear span of the set
$\{\varphi_{k,m}, \,k,m\geq0\}$. In this way we by-pass the problems
raised in \cite{bag1}, and we can avoid replacing the hamiltonian
(\ref{31}) with the approximated hamiltonian (\ref{35}); (ii) we get
an extra bonus which suggests that (\ref{app2}) and (\ref{app3}) are
good definitions: indeed we find that
$$(c^\dagger)^{\,\hat P}=(c^{\,\hat P})^\dagger,$$ and we omit the
proof of this claim here.

More relevant for us is to deduce some commutation rules which
involve the operators $(c^\sharp)^{\,\hat P}$, where $c^\sharp$ can
be $c$ or $c^\dagger$. We claim that \be  \left\{
\begin{array}{ll}\vspace*{2mm}
\left[{\hat P},c^{\,\hat P}\right]=\left[{\hat
P},{c^\dagger}^{\,\hat P}\right]=0,
\\\vspace*{2mm}
\left[{\hat k},c^{\,\hat P}\right]=\left[c^\dagger\,c,c^{\,\hat
P}\right]=-{\hat P}\,c^{\,\hat P}=-c^{\,\hat P}\,{\hat P}\\
 \left[{\hat k},{c^\dagger}^{\,\hat P}\right]={\hat
P}\,{c^\dagger}^{\,\hat
P}={c^\dagger}^{\,\hat P}\,{\hat P}\\
\end{array}
\right. \label{app4}\en Again, we omit the proof of these rules here
since they can be easily deduced applying both sides of each line
above to a vector $\varphi_{k,m}$ of our orthonormal basis. We
simply remark that, for instance, $[{\hat k},c^{\,\hat P}]=-{\hat
P}\,c^{\,\hat P}$ is an extended version of $[\hat
k,c^{\,l}]=-l\,c^{\,l}$, while $[{\hat k},{c^\dagger}^{\,\hat
P}]={\hat P}\,{c^\dagger}^{\,\hat P}$  extends $[\hat
k,{c^\dagger}^{\,l}]=l\,{c^\dagger}^{\,l}$.

\appendix
\renewcommand{\theequation}{B.\arabic{equation}}

% Section 1

 \section{\hspace{-14.5mm} Appendix B:  Few results on the stochastic limit}

In this Appendix we will briefly summarize some of the basic facts
and properties concerning the SLA which are used in Section IV. We
refer to \cite{accbook} and references therein for more details.

Given an open system ${\cal S}+{\cal R}$ we write its hamiltonian
$H$ as the sum of two contributions, the free part $H_0$ and the
interaction $\lambda H_I$. Here $\lambda$ is a coupling constant,
$H_0$ contains the free evolution of both the system ${\cal S}$ and
the reservoir ${\cal R}$, while $H_I$ contains the interaction
between ${\cal S}$ and ${\cal R}$. Working in the interaction
picture, we define $H_I(t)=e^{iH_0t}H_Ie^{-iH_0t}$ and the so called
wave operator $U_\lambda(t)$ which is the solution of the following
differential equation \be
\partial_t U_\lambda(t)=-i\lambda H_I(t)U_\lambda(t),
\label{a1} \en with the initial condition $U_\lambda(0)=\1$. Using
the van-Hove rescaling $t\rightarrow \frac{t}{\lambda^2}$, see
\cite{accbook} for instance, we can rewrite the same equation in a
form which is more convenient for our perturbative approach, that is
\be
\partial_t U_\lambda\left(\frac{t}{\lambda^2}\right)=-\frac{i}{\lambda}
H_I\left(\frac{t}{\lambda^2}\right)U_\lambda\left(\frac{t}{\lambda^2}\right),
\label{a2} \en with the same initial condition as before. Its
integral counterpart is \be
U_\lambda\left(\frac{t}{\lambda^2}\right)=\1-\frac{i}{\lambda}
\int_0^t
H_I\left(\frac{t'}{\lambda^2}\right)U_\lambda\left(\frac{t'}{\lambda^2}\right)dt',
\label{a3} \en which is the starting point for a perturbative
expansion, which works in the following way.

We will limit ourself here to consider the zero temperature
situation. Then let $\varphi_0$ be the ground vector of the
reservoir and $\xi$ a generic vector of the system. Now we put
$\varphi_0^{(\xi)}=\varphi_0\otimes\xi$. We want to compute the
limit, for $\lambda$ going to $0$, of the first non trivial order of
the mean value of the perturbative expansion of
$U_\lambda(t/\lambda^2)$ above in $\varphi_0^{(\xi)}$, that is the
limit of \be I_\lambda(t)=\left(-\frac{i}{\lambda}\right)^2\int_0^t
dt_1 \int_0^{t_1}dt_2\langle
H_I\left(\frac{t_1}{\lambda^2}\right)H_I\left(\frac{t_2}{\lambda^2}\right)\rangle_{\varphi_0^{(\xi)}},
\label{a4} \en for $\lambda\rightarrow 0$. Under some regularity
conditions on the functions which are used to smear out the
(typically) bosonic fields of the reservoir, this limit is shown to
exist for many relevant physical models, see \cite{accbook} and
\cite{bag2}.  We define $I(t)=\lim_{\lambda\rightarrow
0}I_\lambda(t)$. In the same sense of the convergence of the
(rescaled) wave operator $U_\lambda(\frac{t}{\lambda^2})$ (the
convergence in the sense of correlators), it is possible to check
that also the (rescaled) reservoir operators converge and define new
operators which do not satisfy canonical commutation relations but a
modified version of these, \cite{bag2}.  Moreover, these limiting
operators depend explicitly on time and they live in a Hilbert space
which is different from the original one. In particular, they
annihilate a vacuum vector, $\eta_0$,
 which is no longer the original one, $\varphi_0$.

It is not difficult to deduce the form of a time dependent
self-adjoint operator $H_I^{(sl)}(t)$, which depends on the system
operators and on the limiting operators of the reservoir, such that
the  first non trivial order of the mean value of the expansion of
$U_t=\1-i\int_0^tH_I^{(sl)}(t')U_{t'}dt'$ on the state
$\eta_0^{(\xi)}=\eta_0\otimes\xi$ coincides with $I(t)$. The
operator $U_t$  defined by this integral equation is called again
the {\em wave operator}.

The form of the generator follows now from an operation of normal
ordering. More in details, we start defining the flux of an
observable  $\tilde X=X\otimes \1_{r}$, where $\1_{r}$ is the
identity of the reservoir and $X$ is an observable of the system, as
$j_t(\tilde X)=U_t^\dagger \tilde XU_t$. Then, using the equation of
motion for $U_t$ and $U_t^\dagger$, we find that $\partial_t
j_t(\tilde X)=iU_t^\dagger [H_I^{(sl)}(t),\tilde X]U_t$. In order to
compute the mean value of this equation on the state
$\eta_0^{(\xi)}$, so to get rid of the reservoir operators, it is
convenient to compute first the commutation relations between $U_t$
and the limiting operators of the reservoir. At this stage the so
called {\em time consecutive principle} is used in a very heavy way
to simplify the computation. This principle, which has been checked
for many classes of physical models, \cite{accbook}, states that, if
$\beta(t)$ is any of these limiting operators of the reservoir, then
\be [\beta(t),U_{t'}]=0, \mbox{ for all } t>t'. \label{a5} \en Using
this principle and recalling that $\eta_0$ is annihilated by the
limiting annihilation operators of the reservoir, it is now a simple
exercise to compute $\langle\partial_t j_t(\tilde
X)\rangle_{\eta_0^{(\xi)}}$ and, by means of the equation
$\langle\partial_t j_t(\tilde X)\rangle_{\eta_0^{(\xi)}}=\langle
j_t(L(\tilde X))\rangle_{\eta_0^{(\xi)}}$, to identify the form of
the generator of the physical system, which allows us to obtain
equations of motion in general much easier than the original ones,
since the reservoir {\em disappear}.

\end{document}